\documentclass{article}
\usepackage{amssymb}
\usepackage{amsmath}
\usepackage{amsthm}
\usepackage{color}
\usepackage[colorlinks]{hyperref}
\usepackage[a4paper]{geometry}
\usepackage{algorithm}
\usepackage{algpseudocode}
\usepackage{array}
\usepackage[caption=false,font=normalsize,labelfont=sf,textfont=sf]{subfig}
\usepackage{textcomp}
\usepackage{stfloats}
\usepackage{url}
\usepackage{verbatim}
\usepackage{graphicx}
\usepackage{cite}
\usepackage{amsthm}
\usepackage{mathabx}
\usepackage{hyperref}
\usepackage{caption}
\usepackage{tablefootnote}
\usepackage{pgf,tikz,pgfplots}
\usepackage{mathrsfs}
\usepackage{bm}
\usepackage{xpatch}
\usepackage{xcolor}
\usetikzlibrary{arrows}
\usepackage{multirow}
\usepackage{makecell}

\newtheorem{theorem}{Theorem}[section]
\newtheorem{lemma}[theorem]{Lemma}
\newtheorem{prop}[theorem]{Proposition}
\newtheorem{cor}[theorem]{Corollary}

\theoremstyle{definition}
\newtheorem{defn}{Definition}[section]
\newtheorem{example}{Example}[section]

\theoremstyle{remark}
\newtheorem{remark}{Remark}

\newcommand{\C}{\mbox{$\cal C$}}

\newcommand{\ben}{\begin{equation*}}
\newcommand{\een}{\end{equation*}}

\newcommand{\F}{\mathbb{F}}

\newcommand{\Hull}{\operatorname{Hull}}
\newcommand{\rank}{\operatorname{rank}}

\newcommand{\wt}{\operatorname{wt}}

\def\CC{{\mathcal C}}
\def\x{{\mathbf x}}
\def\y{{\mathbf y}}

\title{Shortest self-orthogonal embeddings of binary linear codes}
\author{Junmin An\thanks{junmin0518@sogang.ac.kr, Department of Mathematics and Institute for Mathematical and Data Sciences, Sogang University, Seoul, Korea}, 
Nathan Kaplan\thanks{nckaplan@math.uci.edu, Department of Mathematics, University of California, Irvine, USA}
Jon-Lark Kim\thanks{jlkim@sogang.ac.kr, Department of Mathematics and Institute for Mathematical and Data Sciences, Sogang University, Seoul, Korea},
Jinquan Luo\thanks{luojinquan@ccnu.edu.cn, School of Mathematics and Statistics \& Hubei Key Laboratory of Mathematical Sciences, Central China Normal University, Wuhan, China},
Guodong Wang\thanks{wanggdmath@163.com, School of Mathematics and Statistics \& Hubei Key Laboratory of Mathematical Sciences, Central China Normal University, Wuhan, China}}
\date{}

\begin{document}
\maketitle

\begin{abstract}
There has been recent interest in the study of shortest self-orthogonal embeddings of binary linear codes, since many such codes are distance-optimal self-orthogonal codes.  Several authors have studied the length of a shortest self-orthogonal embedding of a given binary code $\C$, or equivalently, the minimum number of columns that must be added to a generator matrix of $\C$ to form a generator matrix of a self-orthogonal code.   In this paper, we use properties of the hull of a linear code to determine the length of a shortest self-orthogonal embedding of any binary linear code. We focus on Hamming codes and Reed-Muller codes.  We show that a shortest self-orthogonal embedding of a binary Hamming code is self-dual, and propose two algorithms to construct self-dual codes from Hamming codes $\mathcal H_r$. Using these algorithms, we construct a self-dual $[22, 11, 6]$ code, called the shortened Golay code, from the binary $[15, 11, 3]$ Hamming code $\mathcal H_4$, and construct a self-dual $[52, 26, 8]$ code from the binary $[31, 26, 3]$ Hamming code $\mathcal H_5$. We use shortest self-orthogonal embeddings of linear codes to obtain many inequivalent distance-optimal self-orthogonal codes of dimensions $7$ and $8$ for several lengths. Four of the codes of dimension $8$ that we construct are codes with new parameters, namely $[91, 8, 42],\, [98, 8, 46],\,[114, 8, 54]$, and $[191, 8, 94]$.\\

\end{abstract}

\section{Introduction}
Since the beginning of coding theory, the dual of a linear code has been a central topic of study. A code is self-orthogonal if it is a subset of its own dual.  The highly interesting combinatorial properties of this special class of codes have led to applications in various other fields including $t$-designs~\cite{Hadara2004, Bachoc2004}, lattices~\cite{Harada2009, Bouyuklieva2013}, and graphs~\cite{Fellah2024}. Moreover, self-orthogonal codes are a key ingredient in the construction of quantum codes~\cite{Galindo2019, Jin2010, Jin2012, Kim2002:2, Lisoneks2014}, because of the important role they play in the CSS construction. Calderbank et al.~\cite{Calderbank1998} transformed finding quantum error-correcting codes into the problem of finding binary self-orthogonal codes. Also, many codes with good parameters are known to be self-orthogonal~\cite{Zhang2025}. Indeed, some codes in the Best Known Linear Codes (BKLC) database in MAGMA~\cite{Magma1994} are self-orthogonal. For all of these reasons, constructions and characterizations of self-orthogonal codes have been continuously studied.

In this line of research, Pless~\cite{Pless1968} enumerated the number of self-orthogonal codes over a finite field $\mathbb{F}_q$ in terms of finite geometry, and using this result, she proved the uniqueness of the Golay codes. After this foundational work, enumeration and characterization of self-orthogonal codes have been studied over various rings such as $\mathbb{Z}_4$, $\mathbb{Z}_{p^2}$, Galois rings and finite chain rings~\cite{Betty2009, Leon1982, Nagata2009, Nagata2013, Vasquez2019, Yadav2021}. In addition to enumeration, efforts were also focused on the classification of self-orthogonal codes. Pless~\cite{Pless1972} and Pless and Sloane~\cite{Pless1975} classified self-orthogonal and self-dual codes, respectively, for certain lengths and dimensions. Bouyukliev et al.~\cite{Bouyukliev2006} provided a classification of distance-optimal self-orthogonal codes of length less than or equal to $40$ using computational methods. Using systems of linear equations, Li et al.~\cite{Li2008} characterized distance-optimal self-orthogonal codes of dimension $4$. Beyond binary codes, the classification of self-orthogonal codes has been extended to larger fields and rings such as $\mathbb{F}_3$, $\mathbb{F}_4$, $\mathbb{Z}_4$ and $\mathbb{F}_2+u\mathbb{F}_2$. An overview of these results on classification and characterization of self-orthogonal codes can be found in the survey paper by Huffman~\cite{Huffman2005}.

Recently, Kim et al.~\cite{Kim2021} devised an algorithm to construct distance-optimal self-orthogonal codes by adding columns to the generator matrix of a linear code $\mathcal{C}$, which they call the self-orthogonal embedding method. They proved that at most three columns are needed to embed a linear code of dimension three into a self-orthogonal code, and at most five columns are required to embed a linear code of dimension four into a self-orthogonal code. However, the algorithm they suggested can only be applied to linear codes with dimensions at most four. Kim and Choi~\cite{Kim2022} extended this result, providing an algorithm to find a shortest self-orthogonal embedding of a linear code of dimension five or six by using a special matrix called a self-orthogonality matrix. They also proved that the minimum number of columns to be added to embed an $[n, k]$ linear code into a self-orthogonal code is at most $k$ if $k$ is odd, and at most $k+1$ if $k$ is even.

In this paper, we extend the results of~\cite{Kim2022} by investigating the exact value of the length of a shortest self-orthogonal embedding of $\mathcal{C}$. The hull of a linear code $\mathcal{C}$, denoted $\Hull(\mathcal{C})$, is the intersection of $\mathcal{C}$ and its dual, $\mathcal{C}^\perp$.  Let $\ell = \dim(\Hull(\mathcal{C}))$.  We show that if $\mathcal{C}$ is an even code which is not self-orthogonal, then the length of a shortest self-orthogonal embedding of $\mathcal{C}$ is $n+k-\ell+1$. Otherwise, the length of a shortest self-orthogonal embedding of $\mathcal{C}$ is $n+k-\ell$. We find the lengths of shortest self-orthogonal embeddings of binary Reed-Muller codes and binary Hamming codes. We also propose algorithms to find the shortest self-orthogonal embedding of a linear code. With this algorithm, we find all shortest self-orthogonal embeddings of the binary Hamming code $\mathcal{H}_4$ up to equivalence. Among these embeddings, we show that one corresponds to the shortened Golay code, which provides an interesting connection between two important classes of codes, the Hamming codes and Golay codes. Finally, using shortest self-orthogonal embeddings, we construct many inequivalent distance-optimal self-orthogonal codes of dimension $7$ with lengths up to $126$ and of dimension $8$ with lengths up to $254$, respectively. We also obtained several codes of dimension $8$ with new, previously unknown parameters including $[91, 8, 42],\, [98, 8, 46],\,[114, 8, 54]$, and $[191, 8, 94]$.

This paper is organized as follows. Section 2 provides some preliminaries on coding theory and introduces previous results on self-orthogonal embeddings. In Section 3, we present our main results on the length of a shortest self-orthogonal embedding of a binary linear code. We consider separate cases based on whether $k-\ell$ is even or odd. We also provide the exact length of shortest self-orthogonal embeddings of Reed-Muller codes and Hamming codes. Section 4 describes an algorithm to embed Hamming codes into self-orthogonal codes. In Section 5, we apply our results about shortest self-orthogonal embeddings to construct distance-optimal self-orthogonal codes of dimensions $7$ and $8$, identifying several codes with new parameters. We conclude our paper in Section 6.

\section{Preliminaries}\label{sec:prelim}
For a general reference on coding theory, we recommend \cite{Pless1998, Macwilliams1977, Huffman2003}.

Let $\mathbb{F}_2$ be the finite field with two elements.
A subspace $\CC$ of $\mathbb{F}_2^n$ with dimension $k$ is called a {\it linear code} of length $n$ and dimension $k$, and we say that $\C$ is an $[n, k]$ code.
The elements of $\CC$ are called {\it codewords}. A {\it generator matrix} for $\CC$, denoted by $G(\CC)$, is a $k\times n$ matrix whose rows form a basis for $\CC$. For a $k\times n$ matrix $G$, we denote the linear code generated by the rows of $G$ by $\langle G\rangle$. Two codes $\mathcal{C}_1$ and $\mathcal{C}_2$ are said to be {\it (permutation) equivalent} if there exists a permutation of columns $\sigma$ such that $\sigma\mathcal{C}_1=\mathcal{C}_2$.

For $\x = (x_1,x_2,\ldots,x_n),\y = (y_1,y_2,\ldots,y_n) \in \mathbb{F}_2^n$, define 
\[
\x \cdot \y := \sum_{i=1}^n x_iy_i.
\]
For a linear code $\CC$, the {\it dual} of $\CC$ is defined by
\[
\CC^{\perp} := \left\{ \x \in \mathbb{F}_2^n \; \middle| \; \x \cdot \y =0\: {\mbox{for  all }} \y \in \CC\right\}.
\]
A linear code $\CC$ satisfying $\CC \subseteq \CC^\perp$ (resp. $\CC = \CC^\perp$) is called {\it self-orthogonal} (resp. {\it self-dual}). The {\it hull} $\operatorname{Hull}(\mathcal{C})$ of a linear code $\mathcal{C}$ is defined as the intersection of $\mathcal{C}$ and its dual $\mathcal{C}^\perp$. If $\operatorname{Hull}(\mathcal{C})=\{0\}$, then $\mathcal{C}$ is called {\it linear complementary dual} (LCD). Moreover, if $\operatorname{Hull}(\mathcal{C})=\mathcal{C}$, then $\mathcal{C}$ is self-orthogonal. The dimension of the hull of a code will be a major tool in our arguments. We briefly recall how to determine this dimension in terms of the rank of a certain matrix.
\begin{prop}{\rm(\cite[Proposition 3.1]{GueJitGul})} \label{prop-hull}
Let $\mathcal{C} \subseteq \F_2^n$ be a linear code of dimension $k$ and let $G$ be a generator matrix for $\C$.  Then
\[
\rank(G G^T) = k - \dim(\Hull(\C)).
\]
In particular, $\C$ is self-orthogonal if and only if $GG^T= \mathcal{O}$ where $\mathcal{C}$ denotes the zero matrix of a proper size, and $\C$ is LCD if and only if $GG^T$ is invertible.
\end{prop}

Throughout this paper, we write $\mathcal{O}$ to denote the zero matrix, whose size is usually clear from context. We write ${\bf{1}}$ to denote the all-one vector and ${\bf{0}}$ to denote the all-zero vector.

For $\x\in \mathbb{F}_2^n$, the {\it (Hamming) weight} ${\rm wt}(\x)$ is the number of nonzero coordinates in $\x$. For $\x,\y \in \mathbb{F}_2^n$, the {\it (Hamming) distance} $d(\x,\y)$ between $\x$ and $\y$ is defined as 
\[
d(\x, \y)={\rm wt}(\x-\y).
\]
A linear code $\C$ is called an {\it even code} if the weight of every codeword in $\C$ is even. A binary code that is not even is called an \textit{odd code}. If $\mathcal{C}$ is an even LCD code, then the following theorem shows that its dimension must be even.
\begin{theorem}{\rm(\cite[Theorem 5]{CMTQ-2019-1}}\label{LCDbasis}
Let $\mathcal{C}$ be an even code over $\mathbb{F}_2$. Then $\mathcal{C}$ is LCD if and only if the dimension $k$ of $\mathcal{C}$ is even and there exists a basis $\mathbf{c}_1, \mathbf{c}_1', \mathbf{c}_2, \mathbf{c}_2',\ldots, \mathbf{c}_{k/2}, \mathbf{c}_{k/2}'$ of $\mathcal{C}$ such that for any $i, j\in\{1, 2, \ldots, k/2\}$, the following holds:
\begin{enumerate}
    \item[(i)] $\mathbf{c}_i\cdot\mathbf{c}_j=\mathbf{c}_i'\cdot\mathbf{c}_j'=0$,
    \item[(ii)] $\mathbf{c}_i\cdot\mathbf{c}_j'=0$ if $i\ne j$, and
    \item[(iii)] $\mathbf{c}_i\cdot\mathbf{c}_i'=1$.
\end{enumerate}
\end{theorem}

The {\it minimum distance} of $\CC$ is the minimum of the distances between any two distinct codewords.
An {\it $[n,k,d]$ code} $\CC$ is an $[n,k]$ code with minimum distance $d$.
A linear $[n,k, d]$ code $\CC$ is called \textit{distance-optimal} if its minimum distance $d$ is the highest among all $[n,k]$ codes. A self-orthogonal $[n,k, d]$ code with the highest minimum distance among all self-orthogonal $[n, k]$ codes is called an \textit{distance-optimal self-orthogonal code}. For a given length $n$ and dimension $k$, we denote the minimum distance of a diatance-optimal $[n, k]$ self-orthogonal code by $d_{SO}(n, k)$.

In what follows, we give a definition of a self-orthogonal embedding of $\CC$ from~\cite{Kim2021}, which does not depend on a choice of generator matrix for $\C$.

\begin{defn}{\em
	Let $\CC$ be a binary $[n,k]$ code.
		\begin{enumerate}
 	\item[{(i)}]  A \emph{self-orthogonal embedding} of $\CC$ is a self-orthogonal code $\tilde{\C} \subset \mathbb F_2^{n'}$ ($n' \ge n$) such that there exists a subset of coordinates of $\tilde{\C}$ where puncturing $\tilde{\mathcal{C}}$ in those coordinates yields $\mathcal{C}$.
    \item[{(ii)}] A self-orthogonal embedding of $\CC$ is called a \emph{shortest self-orthogonal embedding} of $\CC$ if its length is shortest among all self-orthogonal embeddings of $\CC$.
	\end{enumerate}
}	
\end{defn}

Note that any self-orthogonal embedding $\tilde{\mathcal{C}}$ of $\C$ is permutation equivalent to a self-orthogonal embedding of $\C$ for which puncturing in the last $n'-n$ coordinates gives $\C$. Thus, we have the following.

\begin{lemma}
Let $G$ be a generator matrix for $\C$. Then $\tilde{\mathcal{C}}$ is a self-orthogonal embedding of $\C$ if and only if $\tilde{\mathcal{C}}$ is self-orthogonal and is permutation equivalent to a code with a generator matrix of the form $[G~|~S]$ for some matrix $S$.
\end{lemma}

Therefore, in order to obtain a self-orthogonal embedding of a binary linear code $\C$ with generator matrix $G$, it is enough to add columns to the right of $G$ until we get a self-orthogonal code.

Kim et al.~\cite{Kim2021} showed the following.
\begin{enumerate}
 	\item[{(i)}] Let $\C$ be a binary $[n,2]$ code or a binary $[n,3]$ code.  A shortest self-orthogonal embedding of $\C$ has length at most $n+3$.
	\item[{(ii)}] Let $\C$ be a binary $[n,4]$ code.  A shortest self-orthogonal embedding of $\C$ has length at most $n+5$.
\end{enumerate}

\begin{lemma}{\rm(\cite[Theorem 13]{Kim2022})}\label{lem-Kim2022Thm}
 Let  $\mathcal C$ be any binary $[n, k]$ code.  The length of a shortest self-orthogonal embedding $\tilde{\mathcal{C}}$ of $\mathcal{C}$ is at most
\begin{enumerate}
  \item[{(i)}] $n+k$ if $k$ is odd, or
  \item[{(ii)}] $n+k+1$  if $k$ is even.
\end{enumerate}
\end{lemma}

We first consider shortest self-orthogonal embeddings of $\mathbb F_2^n$.
\begin{prop}
 Let $\mathbb F_2^n$ be the binary code with generator matrix $I_n$.
 \begin{enumerate}
 \item[{(i)}]
 Any shortest self-orthogonal embedding of $\mathbb F_2^n$ is equivalent to a self-dual $[2n, n]$ code with generator matrix $[I_n~|~A]$, where $AA^T=I_n$.

  \item[{(ii)}]
  Conversely, any binary self-dual $[2n, n]$ code is equivalent to a shortest self-orthogonal embedding of $\mathbb F_2^n$.
\end{enumerate}

\end{prop}

\begin{proof}
(i) Let $n+m$ be the length of a shortest self-orthogonal embedding $\mathcal C$ of $\mathbb F_2^n$.  Let $[I_n~|~A]$ be a generator matrix for $\mathcal C$ where $A$ is an $n \times m$ matrix. Since $\C$ is self-orthogonal, we have $n+m \ge 2n$, that is, $m \ge n$.  We choose $A = I_n$ and see that the matrix $[I_n~|~I_n]$ generates a self-dual code. Hence $m=n$. So any shortest self-orthogonal embedding of $\mathbb F_2^n$ is equivalent to a self-dual $[2n, n]$ code whose generator matrix $[I_n~|~A]$ satisfies $AA^T=I_n$.

(ii) Conversely, any binary self-dual $[2n, n]$ code is equivalent to a binary code $\mathcal C'$ with generator matrix $[I_n~|~A]$, where $A$ is an $n \times n$ matrix. Clearly, $\mathcal C'$ is a self-orthogonal embedding of $\mathbb F_2^n$. It is a shortest one since we have just shown that a shortest self-orthogonal embedding has length $2n$.
\end{proof}

Therefore, the relationship between a binary linear code $\C$ and a shortest self-orthogonal embedding $\tilde{\mathcal{C}}$ of $\C$ is a natural generalization of the relation between $\mathbb F_2^n$ and a self-dual $[2n, n]$ code. 

\section{Shortest self-orthogonal embeddings of linear codes}\label{sec-main}

This section is organized as follows. Lemma~\ref{lem-length-equal} first reduces the embedding problem to an LCD complement of the hull, while Lemma~\ref{lem-dim-hul} controls the change in the hull dimension caused by appending a single column. These lemmas yield Proposition~\ref{thm-hull}, which restricts the length of a shortest self-orthogonal embedding to at most two possible values. Theorem~\ref{thm-even-odd-even} then determines the exact value according to whether the code is even or odd. Theorems~\ref{embstruct} and~\ref{ortho-gene} establish the structure of any shortest self-orthogonal embedding and parametrize all shortest embeddings by orthogonal matrices, respectively.

\begin{lemma} \label{lem-length-equal}
Suppose that $\mathcal C$ is an $[n, k]$ binary code with
$\ell =\dim({\rm{Hull}}(\mathcal C))$. Then the following hold.

\begin{enumerate}
\item[{(i)}] $\mathcal C$ has a generator matrix of the form
\[
\begin{bmatrix}
G(\mathcal {\rm{Hull}}(\mathcal C))  \\
A 
\end{bmatrix}
\]
where $\langle A\rangle$ is an LCD code.
\item[{(ii)}] The length of a shortest self-orthogonal embedding of $\mathcal C$ equals the length of a shortest self-orthogonal embedding of the code $\langle A\rangle$.

\end{enumerate}
\end{lemma}

\begin{proof}
We first prove (i). Since a basis of $\operatorname{Hull}(\mathcal{C})$ can be extended to a basis of $\mathcal{C}$, $\mathcal{C}$ has a generator matrix
\[
G=\begin{bmatrix}
G(\mathcal {\rm{Hull}}(\mathcal C))  \\
A 
\end{bmatrix}.
\]
Since
\[
k-\ell = \operatorname{rank}(GG^T)=\operatorname{rank}\left(\begin{bmatrix}
G(\mathcal {\rm{Hull}}(\mathcal C))  \\
A 
\end{bmatrix}\begin{bmatrix}
G(\mathcal {\rm{Hull}}(\mathcal C))  \\
A 
\end{bmatrix}^T\right)
=\operatorname{rank}\left(\begin{bmatrix}
    \mathcal{O} & \mathcal{O}\\\mathcal{O} & AA^T
\end{bmatrix}\right)=\operatorname{rank}(AA^T),
\]
$A$ generates an LCD code.

We now prove (ii). Let $\mathcal C'$ be a shortest self-orthogonal embedding of $\mathcal C$. Then $\mathcal C'$ has a generator matrix of the form
\[
G' = \begin{bmatrix}
  G(\mathcal {\operatorname{Hull}}(\mathcal C)) & B \\
A  & D\\
\end{bmatrix}.
\]
Since $\langle[A ~ D]\rangle$ is a subcode of a self-orthogonal code $\mathcal C'$,
  $\langle[A ~ D]\rangle$ is self-orthogonal.  So the length of a shortest self-orthogonal embedding of $\langle A \rangle$ is at most the length of the shortest self-orthogonal embedding of $\mathcal C$.

Suppose $D$ is a matrix such that $\langle [A~ D] \rangle$ is self-orthogonal. Then
\[
G'' =\begin{bmatrix}
 G(\mathcal {\operatorname{Hull}}(\mathcal C)) & \mathcal{O} \\
A  & D\\
\end{bmatrix}
\]
is a generator matrix of a self-orthogonal embedding of $\mathcal C$.  We see that the length of a shortest self-orthogonal embedding of $\mathcal C$ is at most the length of a shortest self-orthogonal embedding of $\langle A \rangle$.  Therefore, the length of a shortest self-orthogonal embedding of $\mathcal C$ equals the length of a shortest self-orthogonal embedding of $\langle A\rangle$.
\end{proof}

\begin{lemma} \label{lem-dim-hul}
Let $G$ be a $k \times n$ matrix over a field and ${\bf{v}}$ be a $k \times 1$ matrix. Let $G_1 = [G ~|~{\bf{v}}]$ be the matrix obtained by appending $\mathbf{v}$ to $G$. Then
\[
{\rm{rank}}(GG^T) -1 \le {\rm{rank}}(G_1 G_1^T)  \le {\rm{rank}}(GG^T) +1.
\]
Let $\mathcal C$ (resp. $\mathcal C_1$) be a linear code with generator matrix $G$ (resp. $G_1$).
Let $\ell$ be the dimension of $\rm{Hull}(\mathcal C)$ and $\ell_1$ be the dimension of $\rm{Hull}(\mathcal C_1)$. Then
\[
\ell -1 \le \ell_1  \le \ell +1.
\]
\end{lemma}

\begin{proof}
It is easy to see that $G_1 G_1^T = G G^T + {\bf{v}}{\bf{v}}^T$. If ${\bf v} = {\bf 0}$, the conclusion is trivial, so suppose ${\bf v} \neq {\bf 0}$.

Note that for any two matrices $A$ and $B$ of the same size over a field, ${\rm{rank}}(A+B) \le  {\rm{rank}}(A) + {\rm{rank}}(B)$. Let $A=GG^T$ and $B={\bf{v}}{\bf{v}}^T$.  We have ${\rm{rank}}(G_1 G_1^T)  \le {\rm{rank}}(GG^T) +1$ since $\rank({\bf{v}}{\bf{v}}^T) = 1$. Similarly, letting $A=GG^T + {\bf{v}}{\bf{v}}^T$ and $B =-{\bf{v}}{\bf{v}}^T$, we have
\[
{\rm{rank}}(GG^T)= {\rm{rank}}(A+B) \le  {\rm{rank}}(A) + {\rm{rank}}(B) = {\rm{rank}}(GG^T + {\bf{v}}{\bf{v}}^T)  + 1.
\]
Therefore,
${\rm{rank}}(G_1G_1^T) \ge {\rm{rank}}(GG^T) -1$. This completes the first part of the proof.

Applying Proposition \ref{prop-hull} proves the second statement.
\end{proof}

We improve Lemma~\ref{lem-Kim2022Thm} by using information about the dimension of the hull of a linear code.

\begin{prop} \label{thm-hull}
Let $\mathcal C$ be a binary $[n, k]$ code with generator matrix $G(\mathcal C)$ and let $\ell =\dim({\rm{Hull}}(\mathcal C))$. Then the smallest number of columns to be added to $G(\mathcal C)$ for the embedding of $\mathcal C$ into a self-orthogonal $[n', k]$ code $\tilde{\mathcal{C}}$ is
\[
\begin{cases}
 k - \ell & {\mbox{ if }} k - \ell {\mbox{ is odd, }} \\
 k - \ell {\mbox{ or }} k- \ell +1 & {\mbox{ if }} k - \ell {\mbox{ is even. }}
\end{cases}
\]
\end{prop}

\begin{proof}
Let us start from $G(\mathcal C)$ and add a column ${\bf{v}}$ to it to get $G_1 = [G(\mathcal C) ~|~{\bf{v}}]$. Let $\C_1 = \langle G_1 \rangle$ and let $\ell_1 = \dim(\Hull(\C_1))$.  Lemma \ref{lem-dim-hul} implies that $\ell_1 \le \ell+1$.  We repeat this process, adding $i$ columns to $G(\mathcal{C})$ to get $G_i$.  Let $\C_i = \langle G_i \rangle$ and let $\ell_i = \dim(\Hull(\C_i))$.  We know that $\ell_i \le \ell + i$.  If $\C_i$ is self-orthogonal, we must have $\ell_i = k$, so we see that the length of a shortest self-orthogonal embedding of $\C$ is at least $n+k-\ell$.

On the other hand, we start from a generator matrix for $\C$ of the form
\[
\begin{bmatrix}
G(\mathcal {\operatorname{Hull}}(\mathcal C)) \\ A
\end{bmatrix}.
\]
Let $\langle[ A ~ B]\rangle$ be a shortest self-orthogonal embedding of $\langle A \rangle$.  Since $\langle A\rangle$ has dimension $k-\ell$, Lemmas~\ref{lem-Kim2022Thm} and~\ref{lem-length-equal} (ii) imply that the number of columns of $B$ is at most $k-\ell$ if $k-\ell$ is odd, and is at most $k-\ell+1$ when $k-\ell$ is even.  We see that the matrix
\[
\begin{bmatrix}
G(\mathcal {\operatorname{Hull}}(\mathcal C)) & \mathcal{O} \\
A  & B \\
\end{bmatrix}
\]
generates a self-orthogonal embedding $\tilde{\mathcal{C}}$ of $\C$. This completes the proof.
\end{proof}

\begin{theorem} \label{thm-even-odd-even}
Let $\mathcal C$ be a binary $[n, k, d]$ code which is not self-orthogonal, and let $\ell =\dim({\rm{Hull}}(\mathcal C))$.
\begin{enumerate}
    \item[(i)] If $\mathcal{C}$ is even, then the length of a shortest self-orthogonal embedding of $\mathcal{C}$ is $n+k-\ell+1$.
    \item[(ii)] If $\mathcal{C}$ is odd, then the length of a shortest self-orthogonal embedding of $\mathcal{C}$ is $n+k-\ell$.
\end{enumerate}
\end{theorem}

\begin{proof}
Case (a). We first assume that $\mathcal{C}$ is even. By Lemma~\ref{lem-length-equal} (i), $\mathcal C$ has a generator matrix of the form
\[
\begin{bmatrix}
G(\mathcal {\operatorname{Hull}}(\mathcal C))  \\
A  \\
\end{bmatrix}
\]
where $A$ is a $(k -\ell) \times n$ matrix such that $\langle A\rangle$ is LCD. By Theorem~\ref{LCDbasis}, $k-\ell$ must be even. Thus, by Proposition~\ref{thm-hull}, the length of a shortest self-orthogonal embedding of $\mathcal C$ is either $n+k-\ell$ or $n+k-\ell +1$.
We want to show that $\mathcal C$ does not have a self-orthogonal embedding of length $n+k -\ell$. By Lemma~\ref{lem-length-equal} (ii), it is enough to show that $\langle A\rangle$ does not have a self-orthogonal embedding of length $n + k -\ell$.

Since $\langle A\rangle$ is LCD, if $\langle A\rangle$ has a self-orthogonal embedding of length $n+k - \ell$, then Lemma~\ref{lem-dim-hul} implies that there would exist a vector ${\bf{v}} \in \mathbb F_2^{k-\ell}$ such that $\dim(\Hull(\langle[A ~{\bf{v}}]\rangle)) =1$.  We argue by contradiction to prove that there is no such vector.

Suppose that ${\bf{c}}=(c_1, \dots, c_n, c^*) \in {\rm{Hull}}(\langle[A ~{\bf{v}}]\rangle)$.  We have ${\bf{c}}\cdot {\bf{c}} = 0$, so $\wt({\bf{c}})$ is even.  Since $\mathcal C$ is even, $\langle A \rangle$ is even. This implies that $\wt((c_1,\ldots, c_n))$ is even.  We conclude that $c^* =0$. But this implies that $(c_1, \dots, c_n)  \in \Hull(\langle A \rangle)$ and since $\langle A \rangle$ is LCD, $(c_1, \dots, c_n) =(0, \dots, 0)$.  We conclude that $\Hull(\langle[A ~ {\bf{v}}]\rangle)$ is trivial, and so $\langle A \rangle$ does not have a self-orthogonal embedding of length $n+k -\ell$.

Case (b). Next, we assume that $\mathcal{C}$ is odd. Since $\C$ is odd, it has a generator matrix of the form
\[
G(\mathcal C) = \begin{bmatrix}
G({\mbox{{\rm{Hull}}}}(\mathcal C)) \\
{\bf{x}} \\
A_0   \\
\end{bmatrix}
\]
where $\wt({\bf{x}})$ is odd and $\langle A_0\rangle = \mathcal C_0$ is an even code with dimension $k -\ell -1$. If ${\bf{x}} \perp \mathcal C_0$, then $\mathcal{C}_0$ is an LCD code. Then by Theorem~\ref{LCDbasis}, $k-\ell$ must be odd. Thus, we consider the following two cases.

(i) Suppose that $k - \ell$ is odd.  Proposition~\ref{thm-hull} implies that a shortest self-orthogonal embedding of $\C$ has length $n+k-\ell$.

(ii) Suppose that $k - \ell$ is even and that ${\bf{x}}$ is not orthogonal to every vector in $\mathcal C_0$. Let ${\bf y}$ be a row of $A_0$ such that ${\bf x} \cdot {\bf y} \neq 0$.  The matrix
\[
\begin{bmatrix}
G(\operatorname{Hull}(\mathcal{C})) \\
{\bf{x}} \\
A_0   \\
\end{bmatrix}
\]
is row equivalent to a matrix
\[
\begin{bmatrix}
G(\operatorname{Hull}(\mathcal{C})) \\
{\bf{x}} \\
{\bf{y}}\\
B_0
\end{bmatrix}
\]
where every row of $B_0$ is orthogonal to ${\bf x}$.  We construct this matrix by replacing each row ${\bf z} \neq {\bf y}$ of $A_0$ that is not orthogonal to ${\bf x}$ with ${\bf z} + {\bf y}$.  We consider the matrix
\[
\tilde{A_1} = \begin{bmatrix}
{\bf{x}} & 1 \\
{\bf{y}} & 1 \\
B_0   & {\bf{0}}\\
\end{bmatrix}.
\]
Since
\[
\begin{bmatrix}
    \mathbf{x}\\\mathbf{y}\\B_0
\end{bmatrix}
\]
generates an LCD code, by Lemma~\ref{lem-dim-hul}, $\dim(\operatorname{Hull}(\langle\tilde{A_1}\rangle))\le 1$. By construction, $\Hull(\langle\tilde{A_1}\rangle) = \langle ( {\bf{x}}~ 1) \rangle$. Hence the column $(1 ~1~ {\bf{0}})^T$ added to
\[
A' = \begin{bmatrix}
{\bf{x}} \\
{\bf{y}} \\
B_0    \\
\end{bmatrix}
\]
increases the dimension of the hull of $\langle A'\rangle$ by 1. Next, let us consider the submatrix consisting of ${\bf{y}}$ and $B_0$. We aim to embed the code generated by this matrix into a self-orthogonal code of the shortest length by adding exactly $k - \ell -1$ columns.

Let
\[
B_1 = \begin{bmatrix}
{\bf{y}} & 1 \\
B_0   & {\bf{0}}\\
\end{bmatrix}.
\]

 It is clear that the dimension of $\langle B_1 \rangle$ is $k- \ell -1$, which is odd. Furthermore, $\Hull(\langle B_1 \rangle)$ is trivial because we explained above that $\Hull(\langle\tilde{A_1}\rangle) = \langle ( {\bf{x}}~ 1) \rangle$.  Proposition~\ref{thm-hull} now implies that a shortest self-orthogonal embedding of $\langle B_1 \rangle$ has length $n+1 + (k-\ell-1)$.

Altogether, we need exactly $1 + (k -\ell -1) = k - \ell$ columns to embed $\mathcal C$ into a self-orthogonal code of the shortest length.
\end{proof}

We give two examples to illustrate Theorem~\ref{thm-even-odd-even}~(ii).
\begin{example}{\em
Let $\mathcal C$ be the binary distance-optimal $[9, 5, 3]$ code $BKLC(GF(2), 9, 5)$ obtained from MAGMA's BKLC database. We give a generator matrix $G$ for this code for which the first three rows form a basis for $\Hull(\mathcal C)$ (so $\ell = 3$), the fourth row has odd weight, and the fifth row has even weight:
\[
G= \left[ \begin{array}{c}
G({\operatorname{Hull}}(\mathcal C)) \\
\hline
{\bf{x}} \\
\hline
A_0   \\
\end{array}
\right] = \left[
\begin{array}{ccccccccc}
0 &1 &0 &0 &1 &1 &0 &0 &1  \\
0 &0 &1 &0 &1 &1 &0 &1 &0  \\
0 &0 &0 &1 &1 &1 &1 &0 &0  \\
\hline
1 &0 &0 &0 &1 &0 &1 &1 &1  \\
\hline
0 &0 &0 &0 &0 &1 &1 &1 &1  \\
\end{array}
\right].
\]

Using the procedure of the proof of Theorem~\ref{thm-even-odd-even}~(ii), we can construct a self-orthogonal code of length 11, dimension 5, and minimum distance 4 whose generator matrix $G_1$ is given as follows.
\[
G_1 =\left[
\begin{array}{ccccccccc|cc}
0 &1 &0 &0 &1 &1 &0 &0 &1 &0 &0 \\
0 &0 &1 &0 &1 &1 &0 &1 &0 &0 &0 \\
0 &0 &0 &1 &1 &1 &1 &0 &0 &0 &0 \\
\hline
1 &0 &0 &0 &1 &0 &1 &1 &1 &1 &0 \\
\hline
0 &0 &0 &0 &0 &1 &1 &1 &1 &1 &1 \\
\end{array}
\right].
\]
Note that $k - \ell =5-3=2$ is even. Even though ${\bf{x}}=(1 ~ 0 ~ 0 ~ 0 ~ 1 ~ 0 ~ 1 ~ 1 ~ 1)$ is not orthogonal to $\langle A_0 \rangle$, we only need two more columns to embed the code generated by
\[
\begin{bmatrix}
    \mathbf{x}\\A_0
\end{bmatrix}
\]
into a self-orthogonal code.}
We have checked from Grassl's table~\cite{Grassl2026} that this self-orthogonal $[11, 5, 4]$ code is distance-optimal among all $[11, 5]$ codes.
\end{example}

\begin{example}{\em
Let $\mathcal C$ be the binary distance-optimal $[11, 7, 3]$ code $BKLC(GF(2), 11, 7)$ obtained from MAGMA's database. We give a generator matrix $G$ for this code for which the first two rows form a basis for Hull$(\mathcal C)$ (so $\ell =2$), the third row has odd weight, and the fourth through seventh rows have even weight:
\[
G= \left[ \begin{array}{c}
G({\operatorname{Hull}}(\mathcal C)) \\
\hline
{\bf{x}} \\
\hline
A_0   \\
\end{array}
\right] = \left[
\begin{array}{ccccccccccc}
0& 0& 0& 1& 1& 1& 1& 0& 0& 0& 0 \\
0& 0& 0& 0& 0& 0& 0& 1& 1& 1& 1 \\
\hline
1& 0& 0& 0& 0& 0& 0& 0& 0& 1& 1 \\
\hline
1& 1& 0& 0& 0& 0& 0& 0& 1& 1& 0 \\
0& 1& 1& 0& 0& 0& 0& 0& 0& 1& 1 \\
0& 0& 1& 1& 0& 0& 1& 0& 0& 0& 0 \\
0& 0& 0& 1& 1& 0& 0& 0& 0& 1& 1 \\
\end{array}
\right].
\]

Using the procedure of the proof of Theorem~\ref{thm-even-odd-even}~(ii), we construct a self-orthogonal code of length 16, dimension 7, and minimum distance 4 whose generator matrix $G_1$ is given as follows:
\[
G_1 =\left[
\begin{array}{ccccccccccc|ccccc}
0& 0& 0& 1& 1& 1& 1& 0& 0& 0& 0 &0 &0 &0 &0 & 0  \\
0& 0& 0& 0& 0& 0& 0& 1& 1& 1& 1 &0 &0 &0 &0 & 0\\
\hline
1& 0& 0& 0& 0& 0& 0& 0& 0& 1& 1 & 1 & 0& 0& 0&0\\
\hline
1& 0& 1& 1& 0& 0& 1& 0& 0& 1& 1& 1& 1& 0& 0& 0 \\
0& 1& 1& 0& 0& 0& 0& 0& 0& 1& 1& 0& 1& 1& 1& 1 \\
1& 1& 0& 0& 0& 0& 0& 0& 1& 1& 0& 0& 0& 1& 1& 0 \\
0& 1& 1& 1& 1& 0& 0& 0& 0& 0& 0& 0& 0& 1& 0& 1 \\
\end{array}
\right].
\]
The weight distribution of this self-orthogonal code is $A_0 =1, A_4 =6, A_6=32, A_8=50, A_{10} =32, A_{12}=6, A_{16}=1$. It is known~\cite{Bouyukliev2006} that there are twenty self-orthogonal (extremal) $[16,7,4]$ codes. This self-orthogonal code that we have constructed from a diatance-optimal $[11, 7, 3]$ code with the hull dimension $2$ is one of them.
}
\end{example}

\begin{cor}\label{cor-emb-en}
 Let $\mathcal E_n$ be the binary even code of length $n\ge 3$ with parameters $[n, n-1, 2]$. Then the following hold.

\begin{itemize}

\item[{(i)}] If $n$ is even, then $\mathcal E_n$ has a shortest self-orthogonal embedding $\mathcal C'$ whose parameters are $[2n-1, n-1, 4]$.

\item[{(ii)}] If $n$ is odd, then $\mathcal E_n$ has a shortest self-orthogonal embedding $\mathcal C'$ whose parameters are $[2n, n-1, 4]$.
\end{itemize}
\end{cor}

\begin{proof}
 For convenience, we choose the generator matrix $G(\mathcal E_n) = [I_{n-1}~|~ {\bf{1}}]$ where $\mathbf{1}$ is the all-one vector of length $n-1$.

We first prove (i). We have $\mathcal E_n^\perp = \langle{\bf 1}\rangle$.  Since $n$ is even, we have $\Hull(\mathcal E_n) = \langle{\bf 1}\rangle$. Then, by Theorem~\ref{thm-even-odd-even}, a shortest self-orthogonal embedding of $\mathcal{E}_n$ has length $2n-1$. We give an example of a shortest self-orthogonal embedding of $\mathcal{E}_n$.

Let $D=J_{n-1}+I_{n-1}$ where $J_{n-1}$ is the $(n-1)\times (n-1)$ all-one matrix. Any two distinct rows of $D$ are not orthogonal since they share $n-3$ ones in common. Now we add $D$ to the right of $G(\mathcal E_n)$ to make a matrix $[G(\mathcal E_n)~|~ D]$. Since any two distinct rows of $G(\mathcal E_n)$ are not orthogonal, it follows that any two distinct rows of $[G(\mathcal E_n)~|~ D]$ are orthogonal.

We now determine the minimum distance of the resulting code. Let ${\bf g}_i$ and ${\bf d}_i$ be the $i$-th row of $G$ and $D$, respectively. Every weight-$2$ codeword of $\mathcal E_n$ is either a row ${\bf g}_i$ or the sum ${\bf g}_i+{\bf g}_j$ of two distinct
rows. In the former case, the corresponding codeword has weight
\[
\wt({\bf g}_i)+\wt({\bf d}_i)=2+(n-2)=n,
\]
whereas in the latter case it has weight
\[
\wt({\bf g}_i+{\bf g}_j)+\wt({\bf d}_i+{\bf d}_j)=2+2=4.
\]
Since $n\ge4$, the resulting code has minimum distance $4$. Therefore, $[G(\mathcal E_n)~|~ D]$ generates a self-orthogonal code with parameters $[2n-1, n-1, 4]$.

Next, we prove (ii). Since $n$ is odd, $\Hull(\mathcal E_n)$ is trivial. Again, by Theorem~\ref{thm-even-odd-even}, a shortest self-orthogonal embedding of $\mathcal{E}_n$ has length $2n$. It is easy to see that concatenating $G(\mathcal{E}_n)$ with a second copy of itself gives a matrix that generates a shortest self-orthogonal embedding of $G(\mathcal{E}_n)$. Indeed, every nonzero
codeword of this code has the form $({\bf c},{\bf c})$ for some
nonzero ${\bf c}\in\mathcal E_n$, and hence has weight
$2\wt({\bf c})$. Therefore, its minimum distance is four.
\end{proof}

\begin{cor}\label{cor-dual}
Let $\mathcal C$ be a binary $[n, k]$ code.

\begin{itemize}

\item If $\C$ is odd, then a shortest self-orthogonal embedding of $\C$ is self-dual if and only if $\mathcal C^{\perp} \subset \mathcal C$.

\item If $\C$ is even, then a shortest self-orthogonal embedding of $\C$ is self-dual if and only if $\C$ is itself self-dual.
\end{itemize}

\end{cor}

\begin{proof}


A shortest self-orthogonal embedding of $\mathcal C$ is self-dual if and only if it has length $2k$.

Consider the case when $\C$ is odd. By Theorem~\ref{thm-even-odd-even}~(ii), the length of a shortest self-orthogonal embedding of $\C$ is $n + k - \ell$ where $\ell =\dim({\rm{Hull}}(\mathcal C))$. We have $\ell \le \dim(\C^{\perp})=n-k$ with equality if and only if $\operatorname{Hull}(\C)=\C^{\perp}$, or equivalently $\C^{\perp} \subset \C$. Since $n + k - \ell \ge 2k$ with equality if and only if $\ell =n -k$, we see that a shortest self-orthogonal embedding of $\C$ is self-dual if and only if $\C^{\perp} \subset \C$.

Next consider the case when $\C$ is even, but not self-orthogonal. Theorem~\ref{thm-even-odd-even} implies that a shortest self-orthogonal embedding $\tilde{\C}$ of $\C$ has length $n+k -\ell +1$. If $\tilde{\mathcal{C}}$ is self-dual, then $n+k-\ell+1=2k$, i.e., $\ell=n-k+1$. Since $\ell \le \dim(\C^{\perp})=n-k$, it is a contradiction. Thus, if \(\mathcal{C}\) is even, then \(\mathcal{C}\) itself must be self-dual for a shortest self-orthogonal embedding of \(\mathcal{C}\) to be self-dual.
\end{proof}

\begin{prop}
Let $\mathcal{H}_r$ be the binary Hamming $[n=2^r-1, k=n-r, 3]$ code where $r \ge 3$. Then the length of a shortest self-orthogonal embedding of $\mathcal{H}_r$ is $2^{r+1}-2r-2$.
\end{prop}
\begin{proof}
We first recall that $\mathcal H_r$ contains the simplex code $\mathcal S_r$, which is also the dual of  $\mathcal H_r$. Thus, by Corollary~\ref{cor-dual}, a shortest self-orthogonal embedding of $\mathcal{H}_r$ is self-dual.

By Theorem~\ref{thm-even-odd-even}~(ii), the length of a shortest self-orthogonal embedding of $\mathcal{H}_r$ is given as
\[
n+k-r=2n-2r=2^{r+1}-2r-2=2k.
\]
\end{proof}
To get an explicit form of the self-orthogonal embedding, let $G(\mathcal S_r)$ denote a generator matrix for $\mathcal S_r$ and let $A$ be a $(n-2r) \times n$ matrix so that
\[
\begin{bmatrix}
G(\mathcal S_r) \\ A
\end{bmatrix}
\]
is a generator matrix for $\mathcal{H}_r$.  Let $[A ~ B]$ be a generator matrix for a shortest self-orthogonal embedding of $\langle A\rangle$.  Then the matrix
\[
G(\tilde{\mathcal H_r}) = \begin{bmatrix}
G(\mathcal S_r) & \mathcal{O} \\
A  & B \\
\end{bmatrix}
\]
generates a self-orthogonal code with length $n+ (n-2r)=2(n-r)$ and dimension $k=n-r$.

In what follows, we consider self-orthogonal embeddings of binary Reed-Muller codes $\mathcal R(r, m)$. If $r=m$, then $\mathcal R(r, m)$ is the whole space $\mathbb F_2^{2^m}$. If $r=m-1$, then $\mathcal R(r, m)$ is the even code $\mathcal E_{2^m}$. Earlier in this section, we considered these two cases. We note that if
\[
r \le \left\lfloor \frac{m-1}{2} \right\rfloor,
\]
then $\mathcal R(r, m)$ is self-orthogonal.  Therefore, we focus on the case where
\[
\left\lfloor \frac{m-1}{2} \right\rfloor < r \le  m-2.
\]

\begin{prop}
Let $\mathcal R(r, m)$ be a binary Reed-Muller $[n, k]$ code, where $n=2^m$. Suppose that
\[
\left\lfloor \frac{m-1}{2} \right\rfloor < r \le  m-2.
\]
Then the length of a shortest self-orthogonal embedding of $\mathcal R(r, m)$ is $2k+1$.
\end{prop}
\begin{proof}
Note that $\mathcal R(r, m)$ contains its dual $\mathcal R(m-r-1, m)$. Since
\[
\mathcal R(r, m) \subset \mathcal R(m-1, m)=\mathcal E_{2^m},
\]
we see that $\mathcal R(r, m)$ is an even code. Moreover, we know that
\[
k - \ell = k-(n-k) = 2k -n =2k -2^m.
\]
By Theorem~\ref{thm-even-odd-even}~(i), the length of a shortest self-orthogonal embedding of $\mathcal R(r, m)$ is
\[
n+k-\ell+1=2k+1.
\]
\end{proof}

The following theorem shows that a shortest self-orthogonal embedding of a linear code $\C$ has a generator matrix of the form presented in the proof of Lemma~\ref{lem-length-equal} (ii).

\begin{theorem}\label{embstruct}
    Let $\mathcal{C}$ be a binary $[n, k, d]$ code with $\ell=\dim(\Hull(\mathcal{C}))$. Choose a generator matrix of the form
    \[
    G(\mathcal{C})=\begin{bmatrix}
        G(\operatorname{Hull}(\mathcal{C}))\\A
    \end{bmatrix}.
    \]
    Let $\tilde{\mathcal{C}}$ be a shortest self-orthogonal embedding of $\C$ and suppose $\tilde{\mathcal{C}}$ has a generator matrix of the form
    \[
    G' =\begin{bmatrix}
        G(\operatorname{Hull}(\mathcal{C})) & B_H \\A & B_A
    \end{bmatrix}.
    \]
    Then $B_H = \mathcal{O}$.
\end{theorem}
\begin{proof}
    Since $\tilde{\mathcal{C}}$ is self-orthogonal, we see that
    \begin{align*}
    G'G'^T&=\begin{bmatrix}
        G(\operatorname{Hull}(\mathcal{C}))G(\operatorname{Hull}(\mathcal{C}))^T+B_HB_H^T & G(\operatorname{Hull}(\mathcal{C}))A^T+B_HB_A^T\\A(G(\operatorname{Hull}(\mathcal{C}))^T)+B_AB_H^T & AA^T+B_AB_A^T
    \end{bmatrix}\\&=
    \begin{bmatrix}
        B_HB_H^T & B_HB_A^T\\B_AB_H^T & AA^T+B_AB_A^T
    \end{bmatrix}=\mathcal{O}.
    \end{align*}
    Consider the $[n, k-\ell]$ code $\langle A\rangle$. Proposition \ref{prop-hull} implies that
\[
0=\dim(\Hull(\langle A\rangle))=k-\ell-\operatorname{rank}(AA^T),
\]
which shows that $\rank(A A^T) = k-\ell$ and so $\det(AA^T)=1$.

Proposition~\ref{thm-hull} implies that a shortest self-orthogonal embedding of $\C$ has length $n+k-\ell$ or $n+k-\ell+1$.  We consider each case.

    Suppose that the length of $\tilde{\mathcal{C}}$ is $n+k-\ell$. This means that $B_A$ is a square matrix.  Since $AA^T+B_AB_A^T=\mathcal{O}$, we have
    \[
    1=\det(AA^T)=\det(B_AB_A^T)=\det(B_A)^2=\det(B_A).
    \]
    So, $\rank(B_A) = k-\ell$. Since $B_HB_A^T=\mathcal{O}$ and $B_A$ is invertible, we conclude that $B_H=\mathcal{O}$.

    Next, suppose that the length of $\tilde{\mathcal{C}}$ is $n+k-\ell + 1$, that is, $B_A$ is a $(k-\ell)\times (k-\ell+1)$ matrix. Since $\det(B_AB_A^T)=1$, we have
    \[
    k-\ell=\operatorname{rank}(B_AB_A^T)\le \operatorname{rank}(B_A)\le k-\ell,
    \]
    which gives $\operatorname{rank}(B_A)=k-\ell$. Then we have $\dim(\langle B_A\rangle^\perp)=1$.

    Suppose $\mathbf{v}\in \langle B_A\rangle^\perp$ is nonzero and $\mathbf{v}\cdot\mathbf{v}=0$. Let $B'$ be the $(k-\ell+1)\times (k-\ell+1)$ matrix obtained by adjoining $\mathbf{v}$ as the last row of $B_A$, that is,
    \[
    B'=\begin{bmatrix}
        B_A\\\mathbf{v}
    \end{bmatrix}.
    \]
    Then we have
    \[
    B'(B')^T=\begin{bmatrix}
        B_A\\\mathbf{v}
    \end{bmatrix}\begin{bmatrix}
        B_A^T & \mathbf{v}^T
    \end{bmatrix}=
    \begin{bmatrix}
        B_AB_A^T & \mathbf{0}\\\mathbf{0} & 0
    \end{bmatrix}.
    \]
    This gives $0=\det(B'(B')^T)=\det(B')$.  Since $\rank(B_A) = k-\ell$ we see that $\mathbf{v}$ is in the row space of $B_A$, that is, $\mathbf{v}\in \langle B_A\rangle$. So, we have $\mathbf{v}\in \Hull(\langle B_A\rangle)$. Proposition \ref{prop-hull} implies that
    \[
    \dim(\Hull(\langle B_A\rangle))=k-\ell-\operatorname{rank}(B_AB_A^T)=0,
    \]
    so this is a contradiction. We conclude that any nonzero vector orthogonal to every row of $B_A$ must have odd weight. Therefore, we conclude that every row of $B_H$ must be equal to $\mathbf{0}$.
\end{proof}

The following theorem shows that, once one shortest self-orthogonal embedding is known, all other shortest self-orthogonal embeddings can be obtained from it via the action of the orthogonal group.

\begin{theorem}\label{ortho-gene}
    Let $\tilde{\mathcal{C}}$ be a shortest self-orthogonal embedding of a binary $[n, k]$ linear code $\mathcal{C}$ with $\dim(\operatorname{Hull}(\mathcal{C}))=\ell$. Let 
    \[
    \tilde{G} = \begin{bmatrix}
        G(\operatorname{Hull}(\mathcal{C})) & \mathcal{O} \\A & B
    \end{bmatrix}
    \]
    be a generator matrix for $\tilde{\mathcal{C}}$.  

    Let $\mathcal{C}'$ be the code generated by the matrix 
    \[
    G' = \begin{bmatrix}
        G(\operatorname{Hull}(\mathcal{C})) & \mathcal{O} \\A & D
    \end{bmatrix}
    \]
    where $D$ has the same dimensions as $B$.  Then $\mathcal{C}'$ is a shortest self-orthogonal embedding of $\mathcal{C}$ if and only if $D = BR$ for some orthogonal matrix $R$.
\end{theorem}
\begin{proof}
    If $\mathcal{C}$ is already self-orthogonal, then the result is immediate. For the rest of the proof we assume that $\mathcal{C}$ is not self-orthogonal.
    
    We first consider the case where $D = BR$ for an orthogonal matrix $R$. Since
    \[
    G'(G')^T=\begin{bmatrix}
        \mathcal{O} & \mathcal{O}\\\mathcal{O} & AA^T+BR(BR)^T
    \end{bmatrix}=\begin{bmatrix}
        \mathcal{O} & \mathcal{O}\\\mathcal{O} & AA^T+BB^T
    \end{bmatrix}=\mathcal{O},
    \]
    $G'$ generates a shortest self-orthogonal embedding of $\mathcal{C}$.

    Conversely, suppose that $G'$ generates a shortest self-orthogonal embedding of $\mathcal{C}$. Our goal is to show that $D = BR$ for some orthogonal matrix $R$.

    Note that $DD^T=AA^T=BB^T$. Since $\operatorname{rank}(AA^T)=k-\ell$, we have $\operatorname{rank}(B)=\operatorname{rank}(D)=k-\ell$. Let $n+m$ be the length of a shortest self-orthogonal embedding of $\mathcal{C}$. By Proposition~\ref{thm-hull}, $m=k-\ell$ or $m=k-\ell+1$.

    Suppose that $m=k-\ell$. Since both $B$ and $D$ are nonsingular, set $R=B^{-1}D$. Then we have $D=BR$ and
    \[
    RR^T=B^{-1}DD^T(B^{-1})^T=B^{-1}BB^T(B^{-1})^T=I.
    \]

    Suppose next that $m=k-\ell+1$. By Theorem~\ref{thm-even-odd-even}, $\mathcal{C}$ is an even code. Thus, every row of $A$, $B$ and $D$ has even weight. Let $\mathbf{1}$ be the all-one vector of length $m$. Define
    \[
    \hat{B}=\begin{bmatrix}
        B\\\mathbf{1}
    \end{bmatrix}\quad\mbox{and}\quad\hat{D}=\begin{bmatrix}
        D\\\mathbf{1}
    \end{bmatrix}.
    \]
    Since $k-\ell$ is even and $m=k-\ell+1$ is odd, we have $\mathbf{1}\cdot\mathbf{1}=1$. Moreover, $B\mathbf{1}^T=D\mathbf{1}^T=\mathbf{0}$. Note that both $\langle B\rangle$ and $\langle D\rangle$ are even codes. Since $\mathbf{1}\notin\langle B\rangle$ and $\mathbf{1}\notin\langle D\rangle$, both $\hat{B}$ and $\hat{D}$ are nonsingular. Set $R=\hat{B}^{-1}\hat{D}$. Since $\hat{D}=\hat{B}R$, we have $D=BR$ and
    \[
    RR^T=\hat{B}^{-1}\hat{D}\hat{D}^T(\hat{B}^{-1})^T=\hat{B}^{-1}\begin{bmatrix}
        DD^T & \mathbf{0}^T\\\mathbf{0} & 1
    \end{bmatrix}(\hat{B}^{-1})^T=
    \hat{B}^{-1}\begin{bmatrix}
        BB^T & \mathbf{0}^T\\\mathbf{0} & 1
    \end{bmatrix}(\hat{B}^{-1})^T
    =\hat{B}^{-1}\hat{B}\hat{B}^T(\hat{B}^{-1})^T=I.
    \]
    Thus, $R$ is orthogonal in both cases.
\end{proof}

\section{Shortest self-orthogonal embedding of binary Hamming codes}

Corollary~\ref{cor-dual} implies that any shortest self-orthogonal embedding of a binary Hamming code $\mathcal H_r$ of length $2^r -1$ is self-dual.  It is natural to ask about the largest minimum distance that occurs for a shortest self-orthogonal embedding of $\mathcal{H}_r$.  In what follows, we describe several methods for answering this question.

\subsection{Embedding Hamming codes by exhaustive search}\label{ExAlgorithm}

The shortest self-orthogonal embedding of the binary Hamming $[7, 4, 3]$ code $\mathcal H_3$ is the extended Hamming $[8, 4, 4]$ code.

Next let us consider the binary Hamming $[15, 11, 3]$ code $\mathcal H_4$. A shortest self-orthogonal embedding of $\mathcal H_4$ is a self-dual $[22, 11]$ code. The minimum distance of this self-dual code can be four or six. By applying Algorithm~\ref{alg-ex} to fill $B$ exhaustively, we have found exactly one self-dual $[22, 11, 4]$ code, and one extremal self-dual $[22, 11, 6]$ code, up to equivalence.

\begin{algorithm}[!htb]
  \caption{Exhaustive search for self-orthogonal embeddings of a linear code}
  \label{alg-ex}
  \begin{algorithmic}[1]
      \State \textbf{Input}: An $[n,k,d]$ code $\mathcal{C}$ with hull dimension $\ell$
      \State ~~~~~~~~ Let $G = [g_1, \ldots, g_k]^T$ be a generator matrix where the first $\ell$ rows span $\text{Hull}(\mathcal{C})$ 
      \State ~~~~~~~~ Let $m=n'-n$ where $n'$ is the length of a shortest self-orthogonal embedding of $\mathcal{C}$
      \State \textbf{Goal}: Find all inequivalent self-orthogonal embeddings $[G \mid B]$

      \Procedure{SearchSOEmbedding}{}
          \State Precompute $\langle g_i, g_j \rangle$ for all $1 \le i,j \le k$
          \State Initialize the set of self-orthogonal embeddings $\mathcal{L} \gets \emptyset$

          \For{$i = 1$ to $k$}
              \If{$i \le \ell$}
                  \State $\mathcal{V}_i \gets \{\mathbf{0}\}$
              \ElsIf{$i = \ell+1$}
                  \State $\mathcal{V}_i \gets \{(1,\ldots,1,0,\ldots,0) \in \mathbb{F}_2^m : \text{all-one at the front, } {\rm wt}(v) \equiv {\rm wt}(g_i) \pmod{2}\}$
              \Else
                  \State $\mathcal{V}_i \gets \{v \in \mathbb{F}_2^m : {\rm wt}(v) \equiv {\rm wt}(g_i) \pmod{2}\}$
              \EndIf
          \EndFor

          \State \Call{Backtrack}{$1, [\ ]$}
          \State \Return $\mathcal{L}$

          \Procedure{Backtrack}{$i, B$}
              \If{$i > k$}
                  \State Construct $\tilde{G} \gets [G \mid B]$
                  \If{$\tilde{G} \tilde{G}^T = \mathcal{O}_k$}
                      \State $\tilde{\mathcal{C}} \gets$ code generated by $\tilde{G}$
                      \If{$\tilde{\mathcal{C}}$ is not equivalent to any code} in $\mathcal{L}$
                          \State $\mathcal{L} \gets \mathcal{L} \cup \{\tilde{\mathcal{C}}\}$
                      \EndIf
                  \EndIf
                  \State \Return
              \EndIf

              \For{$v \in \mathcal{V}_i$}
                  \If{$\langle b_j, v \rangle = \langle g_j, g_i \rangle$ for all $j < i$}
                      \State \Call{Backtrack}{$i+1, B$ concatenated with $[\mathbf{v}]$}
                  \EndIf
              \EndFor
          \EndProcedure
      \EndProcedure
  \end{algorithmic}
\end{algorithm}

We give an example of a self-dual $[22, 11, 4]$ code $\mathcal C_{22,11,4}$ or $U_{22}$ in the notation of~\cite{Pless1975} as a shortest self-orthogonal embedding of $\mathcal H_4$ as follows:
\[
G(\mathcal C_{22,11,4})
= \left[ \begin{array}{c|c}
G(\mathcal S_4) & \mathcal{O} \\
\hline
A  & B \\
\end{array}
\right]
= \left[ \begin{array}{c|c}
G(\mathcal S_4) & \mathcal{O} \\
\hline
{\bf{x}} & {\bf{x}}_1'\\
\hline
A_0   & A_0' \\
\end{array}
\right] = \left[
\begin{array}{c|c}
1 0 1 0 1 0 1 0 1 0 1 0 1 0 1 & 0 0 0 0 0 0 0\\
0 1 1 0 0 1 1 0 0 1 1 0 0 1 1 & 0 0 0 0 0 0 0\\
0 0 0 1 1 1 1 0 0 0 0 1 1 1 1 & 0 0 0 0 0 0 0\\
0 0 0 0 0 0 0 1 1 1 1 1 1 1 1 & 0 0 0 0 0 0 0\\
\hline
1 0 0 0 0 0 0 0 0 0 0 0 0 1 1 & 1 0 0 0 0 0 0 \\
\hline
1 1 0 0 0 0 0 0 0 0 0 0 1 1 0 & 0 0 0 0 1 1 0 \\
0 1 1 0 0 0 0 0 0 0 0 0 0 1 1 & 0 1 1 0 0 0 0 \\
0 0 1 1 0 0 0 0 0 0 1 0 1 1 1 & 0 0 1 1 0 0 0 \\
0 0 0 1 1 0 0 0 0 0 0 0 0 1 1 & 0 0 0 1 1 0 0 \\
0 0 0 0 1 1 0 0 0 0 0 0 1 1 0 & 1 1 0 0 0 0 0 \\
0 0 0 0 0 1 1 0 0 0 0 0 0 1 1 & 0 0 0 0 0 1 1 \\
\end{array}
\right].
\]
The weight distribution of $\mathcal C_{22,11,4}$ is $A_0 =1, A_4 =4, A_6 =73, A_8=318, A_{10} =628, A_{12}=628, A_{14}=318, A_{16}=73, A_{18}=4, A_{22}=1$.

Furthermore, by choosing ${\bf{x}}_2' =(1, 1, 1, 0, 0, 0, 0)$, we have found the shortened Golay $[22, 11, 6]$ code $\mathcal C_{22,11,6}$, or $G_{22}$ in the notation of~\cite{Pless1975}, as a shortest self-orthogonal embedding of $\mathcal H_4$. Its generator matrix is given as follows:
\[
G(\mathcal C_{22,11,6})
= \left[ \begin{array}{c|c}
G(\mathcal S_4) & \mathcal{O} \\
\hline
{\bf{x}} & {\bf{x}}_2'\\
\hline
A_0   & A_0' \\
\end{array}
\right] = \left[
\begin{array}{c|c}
1 0 1 0 1 0 1 0 1 0 1 0 1 0 1 & 0 0 0 0 0 0 0\\
0 1 1 0 0 1 1 0 0 1 1 0 0 1 1 & 0 0 0 0 0 0 0\\
0 0 0 1 1 1 1 0 0 0 0 1 1 1 1 & 0 0 0 0 0 0 0\\
0 0 0 0 0 0 0 1 1 1 1 1 1 1 1 & 0 0 0 0 0 0 0\\
\hline
1 0 0 0 0 0 0 0 0 0 0 0 0 1 1 & 1 1 1 0 0 0 0 \\
\hline
1 1 0 0 0 0 0 0 0 0 0 0 1 1 0 & 1 1 0 1 1 0 0 \\
0 1 1 0 0 0 0 0 0 0 0 0 0 1 1 & 1 0 1 1 0 1 0 \\
0 0 1 1 0 0 0 0 0 0 1 0 1 1 1 & 1 1 0 0 0 0 0 \\
0 0 0 1 1 0 0 0 0 0 0 0 0 1 1 & 0 1 1 1 1 0 0 \\
0 0 0 0 1 1 0 0 0 0 0 0 1 1 0 & 1 1 1 0 0 1 0 \\
0 0 0 0 0 1 1 0 0 0 0 0 0 1 1 & 1 1 0 1 0 0 1 \\
\end{array}
\right].
\]

We remark that as far as we know, this is the first construction of the binary shortened Golay code from the binary Hamming code of length 15. This serves as an example of how the procedure given here can lead to new constructions of interesting self-orthogonal or self-dual codes.


\subsection{Embedding Hamming codes using orthogonal matrices}

In this section, we use orthogonal matrices to provide an algorithm to determine all of the shortest self-orthogonal embeddings of the binary $[2^r-1, 2^r-1-r,3]$ Hamming codes $\mathcal{H}_r$ where $r\ge 3$ up to equivalence.  Let $k=2^r-r-1$. Choose a systematic generator matrix of $\mathcal{H}_r$ as follows:
\[
G(\mathcal{H}_r)=[I_k~|~A].
\]
Then $ G = [A^T~|~ I_r]$ is a generator matrix of $\mathcal{H}_r^\perp = \mathcal{S}_r$, the simplex code.
Since simplex codes are self-orthogonal, we have
\[
  [A^T~|~ I_r][A^T~|~ I_r]^T=\mathcal{O}
\]
which is equivalent to the condition $A^TA=I_r$. In this case, the row vectors of $\mathbf{a}_1, \ldots, \mathbf{a}_r$ of $A^T$ form an orthonormal basis, that is, $\mathbf{a}_i\cdot\mathbf{a}_j=\delta$ of an odd binary LCD code, which we denote by $C_A=\langle A^T\rangle$.

Our goal is to obtain the $k\times (k-r)$ matrix $B$ such that the matrix
\begin{equation}\label{gen-mat-emb}
G_B=[I_k~|~A~|~B]
\end{equation}
generates a self-dual code. For such a matrix,
\[
  G_BG_B^T=\mathcal{O} \quad\Longleftrightarrow\quad [A~|~B][A~|~B]^T = I_k
  \quad\Longleftrightarrow\quad [A~|~B]^T[A~|~B]= I_k.
\]
This holds if and only if the following three equalities are satisfied:
\begin{align}\label{con-B}
  A^TA=I_r, \ \ \ A^T B=\mathcal{O}, \ \ \ B^T B=I_{k-r}.
\end{align}
These conditions are satisfied when both $C_A$ and its dual are odd LCD codes, and
the column vectors of $B$ form an orthonormal basis of $C_A^{\perp}$.

Since $C_A$ is already an odd LCD code and the dual of an LCD code is also LCD, it remains only to show that $C_A^\perp$ is odd. We do this by showing that $C_A$ does not contain $(1,1,\ldots, 1)$.  Since $C_A$ is a linear code obtained by puncturing the last $r$ coordinates of $\mathcal{H}_r^\perp$, and $\mathcal{H}_r^\perp$ is a constant-weight code with weight $2^{r-1}$, the maximum weight of a codeword in $C_A$ is $2^{r-1}-1$, which is not equal to the code length $2^r-r-1$.  Therefore, $C_A$ does not contain the all-one vector.

Next, we describe a procedure to transform a basis for $C_A^\perp$ into an orthonormal basis.

\textbf{Transform method:}
Let $ C $ be an odd binary LCD code, and let $ S = \{\bm{a}_1, \ldots, \bm{a}_s\} $ be an arbitrary basis of $ C $.

\begin{enumerate}
\item \textbf{Step 1.} There exists $ \bm{b}_1 \in S $ such that $ {\rm wt}(\bm{b}_1) $ is odd. Without loss of generality, assume $ \bm{b}_1 = \bm{a}_1 $. Transform the remaining vectors to be orthogonal to $ \bm{b}_1 $ by:
  \begin{align}\label{equation}
    \bm{a}_i \leftarrow \bm{a}_i + \langle \bm{b}_1, \bm{a}_i \rangle \bm{b}_1, \quad i = 2, \ldots, s.
  \end{align}

\item \textbf{Step 2.1.} If there exists a vector with odd weight among the remaining vectors, select such a vector and call it $ \bm{b}_2$. Use Equation~\eqref{equation} to orthogonalize the remaining vectors against $ \bm{b}_2 $.

\item \textbf{Step 2.2.} If all remaining vectors have even weight, select an arbitrary vector denoted as $ \bm{a}_2$.
    There must exist another vector denoted as $ \bm{a}_3 $ such that $ \langle \bm{a}_2, \bm{a}_3 \rangle = 1 $.
    Define:
\[
  \begin{cases}
    \bm{b}_1 \leftarrow \bm{b}_1 + \bm{a}_2 + \bm{a}_3, \\
    \bm{b}_2 \leftarrow \bm{a}_2 + \bm{b}_1, \\
    \bm{b}_3 \leftarrow \bm{a}_3 + \bm{b}_1, \\
    \bm{a}_i \leftarrow \bm{a}_i + \langle \bm{a}_2, \bm{a}_i \rangle \bm{a}_3 + \langle \bm{a}_3, \bm{a}_i \rangle \bm{a}_2, \quad i = 4, \ldots, s.
  \end{cases}
\]
This is clearly an invertible linear transformation. The vectors $ \bm{b}_1, \bm{b}_2, \bm{b}_3 $ are mutually orthogonal, and $\langle \bm{b}_m, \bm{a}_n \rangle = 0$, for $m=1,2,3$ and $n=4,...,s$.

\item Repeat Steps 2.1-2.2 for the remaining vectors until no vectors remain.
\end{enumerate}

\begin{algorithm}[!htb]
  \caption{Converting a basis of a binary odd LCD code to an orthonormal basis}
  \label{alg-ortho}
  \begin{algorithmic}[1]
      \State \textbf{Input:} A binary odd LCD code $ C $
      \State ~~~~~~~~ Let $ S = \{\bm{a}_1, \ldots, \bm{a}_s\} $ be an arbitrary basis of $ C $
      \State \textbf{Goal}: Transform this basis into an orthonormal basis

      \Procedure{GramSchmidtOrthogonalization}{$S$}
          \State Initialize the list $B$ of selected basis vectors
          \State Initialize the set of remaining vectors $ R \gets S $

          \While{$R \neq \emptyset$}
              \If{There exists $ \bm{b} \in R $ such that $ {\rm wt}(\bm{b}) $ is odd}
                  \State Select $ \bm{b} \in R $ with $ {\rm wt}(\bm{b}) $ being odd
                  \State Denote the first vector satisfying this condition as $ \bm{b}_1 = \bm{b} $
                  \State $ B \gets B \cup \{\bm{b}\} $
                  \State $ R \gets R \setminus \{\bm{b}\} $

                  \For{$\bm{a} \in R$}
                      \State $ \bm{a} \gets \bm{a} + \langle \bm{b}, \bm{a} \rangle \bm{b} $
                  \EndFor
              \Else
                  \State Arbitrarily select $ \bm{a}_2' \in R $, and find $ \bm{a}_3' \in R $ such that $ \langle \bm{a}_2', \bm{a}_3' \rangle = 1 $

                  \State $ \bm{b}_1 \gets \bm{b}_1 + \bm{a}_2' + \bm{a}_3' $ (Update the vector $ \bm{b}_1 $ in $ B $)
                  \State $ \bm{b}_2 \gets \bm{a}_2' + \bm{b}_1 $
                  \State $ \bm{b}_3 \gets \bm{a}_3' + \bm{b}_1 $

                  \State $ B \gets B \cup \{\bm{b}_2, \bm{b}_3\} $
                  \State $ R \gets R \setminus \{\bm{a}_2', \bm{a}_3'\} $

                  \For{$\bm{a} \in R$}
                       \State $\bm{a} \gets \bm{a} + \langle \bm{a}_2', \bm{a} \rangle \bm{a}_3' + \langle \bm{a}_3', \bm{a} \rangle \bm{a}_2' $
                  \EndFor
              \EndIf
          \EndWhile

          \State \Return $B$
      \EndProcedure
  \end{algorithmic}
\end{algorithm}

With Theorem~\ref{ortho-gene}, this algorithm proves the following.
\begin{remark}\label{HammingAlg}
With the notations as above, we have the following.
\begin{enumerate}
    \item[(i)] The code generated by the matrix $G_B$ in \eqref{gen-mat-emb} is a self-dual embedding of $\mathcal{H}_r$ if the columns of $B$ consist of an orthonormal basis for the solution space of $A^TX={\bf 0}$.
    \item [(ii)] Another matrix $G_{B'}$ also generates a  self-dual embedding of $\mathcal{H}_r$  if and only if $B'=BR$ for some orthogonal matrix $R$.
\end{enumerate}
\end{remark}

We further show that Remark~\ref{HammingAlg} is not specific to Hamming codes. An analogous result holds for any binary linear code whose shortest self-orthogonal embedding is self-dual.
\begin{prop}
Let $\mathcal{C}$ be a binary linear code that can be embedded into a self-dual code. Let $G(\mathcal{C})=[I~|~A]$ be a generator matrix for $\mathcal{C}$.  Suppose $B$ is a matrix whose columns give an orthonormal basis for the solution space of $A^TX={\bf 0}$.  Then $G_B=[I~|~A~|~B]$ is a generator matrix for a shortest self-orthogonal embedding of $\mathcal{C}$.
\end{prop}
\begin{proof}
    Let $\mathcal{C}_A$ be the code with generator matrix $A^T$. In the argument preceding Remark~\ref{HammingAlg}, the properties of the Hamming code were used only to show that both $\mathcal{C}_A$ and its dual are odd LCD codes. Therefore, for an arbitrary linear code $\mathcal{C}$ that can be embedded into a self-dual code, it is sufficient to show that this condition holds.

    By Corollary~\ref{cor-dual}, if $\mathcal{C}$ is an even code, then $\mathcal{C}$ itself is a self-dual code, and the statement holds trivially. Thus, we assume that $\mathcal{C}$ is odd, in which case $\mathcal{C}^\perp\subset \mathcal{C}$. Since $\mathcal{C}^\perp$ is self-orthogonal, and $[A^T~I]$ is a generator matrix for $\mathcal{C}^\perp$, we have
    \[
    [A^T~|~I] [A^T~|~I]^T= \mathcal{O},
  \]
  that is, $A^TA=I$. Therefore, $\mathcal{C}_A$ is an odd LCD code, and $\mathcal{C}_A^\perp$ is also an LCD code. It remains to show that $\mathcal{C}_A^\perp$ is an odd code.

  Take $\mathbf{c}\in\mathcal{C}$ such that $\wt(\mathbf{c})$ is odd. There is $\mathbf{x}\in\mathbb{F}_2^k$ such that $\mathbf{x}G(\mathcal{C})=(\mathbf{x}~|~\mathbf{x}A)=\mathbf{c}$. Define $\mathbf{y}=\mathbf{x}A$ and $\mathbf{u}=\mathbf{x}+\mathbf{y}A^T$. Since
  \[
  \mathbf{u}A=\mathbf{x}A+\mathbf{y}A^TA=\mathbf{x}A+\mathbf{y}=\mathbf{x}A+\mathbf{x}A=\mathbf{0},
  \]
  we have $\mathbf{u}\in\mathcal{C}_A^\perp$. We now show that $\wt(\mathbf{u})$ is odd.

  Note that
  \begin{equation}\label{eqt}
  {\rm wt}(\mathbf{u})={\rm wt}(\mathbf{x}+\mathbf{y}A^T)\equiv {\rm wt}(\mathbf{x})+{\rm wt}(\mathbf{y}A^T)\pmod 2.
  \end{equation}
  Let $\mathbf{y}=(y_1, \ldots, y_{n-k})$ and $A_i$ be the $i$\textsuperscript{th} column of $A$. We have
  \[
  {\rm wt}(\mathbf{y}A^T)={\rm wt}\left(\sum_i y_iA_i\right)\equiv\sum_i{\rm wt}(y_iA_i)\pmod 2.
  \]
  Since $A^T A$ is the identity matrix, we see that $A_i\cdot A_i=1$ for every $i$.  This implies $\wt(A_i)$ is odd for every $i$. This gives $\wt(\mathbf{y}A^T)\equiv \wt(\mathbf{y})\pmod 2$. Applying this to~\eqref{eqt}, we see that
  \[
  {\rm wt}(\mathbf{u})\equiv {\rm wt}(\mathbf{x})+{\rm wt}(\mathbf{y}A^T)\equiv {\rm wt}(\mathbf{x})+{\rm wt}(\mathbf{y})={\rm wt}(\mathbf{x})+{\rm wt}(\mathbf{x}A)={\rm wt}(\mathbf{c})\equiv 1\pmod 2.
  \]
  Then $\mathbf{u}$ is a codeword in $\mathcal{C}_A^\perp$ with odd weight. This completes the proof.
\end{proof}

Let $\mathcal{O}(s, q)$ denote the orthogonal group of degree $s$ over the finite field $\mathbb{F}_q$ with $q$ elements. In the case $r=4$, $k=11$, the cardinality of $\mathcal{O}(7, 2)$ is  $9\cdot 2^5\cdot 7!$ (see \cite{Macwilliams1969}). Permuting columns of $B$ will produce equivalent self-dual embeddings of $\mathcal{H}_r$. Let $S_7$ denote the symmetric group of permutations on the columns of $B$. Therefore we only need to consider $R$ in a set of representatives
\[
\mathcal{R}=\{R_1, \cdots, R_s\}
\]
such that the disjoint union $R_1S_7 \sqcup \cdots \sqcup R_sS_7$ is equal to $\mathcal{O}(7, 2)$.  The cardinality of $\mathcal{R}$ is
\[
\frac{|\mathcal{O}(7, 2)|}{|S_7|}=288.
\]
Therefore, it is sufficient to search through $288$ cases to find all of the inequivalent self-dual embeddings of $\mathcal{H}_4$.

\begin{example}
We carry out the strategy described above and see that every shortest self-dual embedding of the $[15, 11, 3]$ Hamming code $\mathcal H_4$ is equivalent to either a self-dual $[22, 11, 4]$ code $U_{22}$ in the notation of~\cite{Pless1975}, or the shortened Golay $[22, 11, 6]$ code.

  The binary Hamming $[15, 11, 3]$ code has a systematic generator matrix $G= \left[ I \ A \right]$, where
  \[
    A= \left[\begin{array}{cccc}
       1& 1& 0& 0 \\
       0 &1& 1& 0 \\
       0& 0& 1& 1 \\
       1& 1& 0& 1 \\
       1& 0& 1& 0 \\
       0& 1& 0& 1 \\
       1& 1& 1& 0 \\
       0& 1& 1& 1 \\
       1& 1& 1& 1 \\
       1& 0& 1& 1 \\
       1& 0& 0& 1 \\
    \end{array}\right].
  \]
 Choose a generator matrix of $\mathcal{C}_A^\perp$ as follows:
  \[ \left[\begin{array}{ccccccccccc}
    1& 0 &0& 0& 0& 0& 0 &1& 0& 1& 0 \\
    0& 1& 0& 0& 0& 0& 0& 0& 1& 0& 1 \\
    0& 0& 1& 0& 0& 0& 0& 1& 1& 1& 0 \\
    0& 0& 0& 1& 0& 0& 0& 0& 1& 1& 1 \\
    0& 0& 0& 0& 1& 0& 0& 1& 1 &1 &1 \\
    0& 0& 0& 0& 0& 1& 0& 1& 0& 1& 1 \\
    0& 0& 0& 0& 0& 0& 1& 1& 0& 0& 1 \\
    \end{array}\right].
  \]
By applying Algorithm~\ref{alg-ortho} to the rows of this matrix, we transform it into an orthonormal basis whose vectors form the rows of $B^T$ as follows:
    \[
      B^T=\left[\begin{array}{ccccccccccc}
      1 &0& 0& 0& 0& 0& 0& 1 &0 &1& 0 \\
      0 &1 &0& 0& 0 &0 &0 &0 &1& 0 &1 \\
      0& 1& 1& 0& 0& 0& 0& 1& 0& 1& 1 \\
      1& 0 &0& 1& 0& 0& 0& 1& 1& 0& 1 \\
      1& 1& 1& 1& 1& 0& 0& 1& 0& 0& 1 \\
      1& 1& 0& 0& 0& 0& 1& 0& 1& 1& 0 \\
      1 &1 &1 &0& 0& 1& 1& 0& 0& 1& 1
    \end{array}\right].
    \]
    The matrix $G_B= \left[ I_{11} ~\ A ~\ B \right]$ generates a self-dual code with parameters $[22, 11, 4]$.

    Furthermore, choosing the orthogonal matrix
    \[
    R=\left[\begin{array}{ccccccc}0 & 1 &1& 0& 0& 1& 0 \\
      0 &1 & 0 & 0 & 1& 1& 0 \\
      1& 1 &1& 0& 1 &0 &1 \\
      0 &0 &0 &1 &0& 0& 0 \\
      0& 1& 0& 0 &0 &1& 1 \\
      1 &0 &1& 0 &1& 1& 1 \\
      1 &1 &0& 0& 0& 1 &0 \\
    \end{array}\right],
    \]
    we compute that $\left[ I_{11} ~\ A ~\ B R \right]$ is a generator matrix of a self-dual code with parameters $[22,11,6]$.
\end{example}

\begin{example}
Consider the $[31, 26, 3]$ binary Hamming code $\mathcal{H}_5$. Applying the procedure above, we see that $\mathcal{H}_5$ has a self-dual embedding that is a $[52, 26, 8]$ code.  This code is near-extremal.  A generator matrix $[I_{26} ~\ A_{26}]$ for this code is given below:
\[ \tiny
A_{26}= \left[\begin{array}{cccccccccccccccccccccccccc}
1 & 0 & 1 & 0 & 0 & 0 & 0 & 1 & 0 & 1 & 0 & 0 & 0 & 0 & 1 & 1 & 1 & 0 & 0 & 0 & 1 & 0 & 0 & 0 & 1 & 0\\
0 & 1 & 0 & 1 & 0 & 1 & 1 & 1 & 1 & 0 & 0 & 0 & 1 & 0 & 0 & 0 & 0 & 1 & 0 & 0 & 0 & 0 & 1 & 1 & 1 & 0\\
0 & 0 & 1 & 0 & 1 & 0 & 0 & 0 & 1 & 1 & 1 & 0 & 0 & 1 & 0 & 0 & 1 & 1 & 0 & 0 & 1 & 0 & 0 & 0 & 0 & 0\\
1 & 0 & 1 & 1 & 0 & 1 & 0 & 1 & 1 & 0 & 1 & 0 & 0 & 0 & 0 & 0 & 1 & 0 & 0 & 1 & 0 & 0 & 1 & 1 & 0 & 0\\
0 & 1 & 0 & 1 & 1 & 0 & 1 & 0 & 0 & 1 & 0 & 0 & 0 & 0 & 1 & 1 & 0 & 0 & 1 & 1 & 0 & 1 & 1 & 1 & 1 & 0\\
1 & 0 & 0 & 0 & 1 & 1 & 0 & 1 & 1 & 1 & 0 & 0 & 0 & 0 & 1 & 1 & 0 & 1 & 0 & 1 & 0 & 1 & 0 & 1 & 1 & 0\\
1 & 1 & 1 & 0 & 0 & 1 & 1 & 0 & 1 & 1 & 0 & 1 & 0 & 0 & 1 & 0 & 0 & 1 & 1 & 1 & 0 & 1 & 1 & 0 & 0 & 1\\
0 & 1 & 1 & 1 & 0 & 1 & 0 & 1 & 1 & 1 & 1 & 0 & 1 & 0 & 0 & 1 & 1 & 0 & 1 & 1 & 1 & 0 & 0 & 0 & 0 & 1\\
0 & 0 & 1 & 1 & 1 & 1 & 0 & 0 & 0 & 1 & 1 & 1 & 1 & 1 & 0 & 1 & 0 & 1 & 0 & 1 & 0 & 1 & 1 & 0 & 1 & 0\\
1 & 0 & 1 & 1 & 1 & 1 & 0 & 1 & 0 & 0 & 0 & 1 & 0 & 0 & 1 & 1 & 0 & 0 & 0 & 1 & 0 & 0 & 0 & 0 & 1 & 0\\
1 & 1 & 1 & 1 & 1 & 1 & 1 & 0 & 0 & 1 & 1 & 1 & 1 & 0 & 1 & 0 & 1 & 0 & 1 & 1 & 1 & 1 & 0 & 1 & 0 & 1\\
1 & 1 & 0 & 1 & 1 & 1 & 0 & 0 & 0 & 0 & 0 & 1 & 0 & 0 & 0 & 0 & 0 & 1 & 0 & 0 & 0 & 1 & 1 & 0 & 1 & 1\\
1 & 1 & 0 & 0 & 1 & 0 & 1 & 1 & 0 & 0 & 1 & 1 & 1 & 1 & 0 & 1 & 1 & 1 & 1 & 0 & 1 & 0 & 0 & 0 & 1 & 0\\
1 & 1 & 0 & 0 & 0 & 1 & 1 & 0 & 1 & 0 & 1 & 0 & 0 & 1 & 1 & 0 & 0 & 1 & 0 & 1 & 0 & 1 & 0 & 1 & 0 & 1\\
0 & 1 & 1 & 0 & 0 & 0 & 0 & 0 & 1 & 1 & 0 & 1 & 0 & 1 & 1 & 0 & 0 & 1 & 1 & 1 & 1 & 0 & 0 & 1 & 0 & 1\\
0 & 0 & 1 & 1 & 0 & 1 & 0 & 0 & 1 & 1 & 1 & 1 & 1 & 0 & 0 & 0 & 0 & 1 & 0 & 1 & 1 & 0 & 0 & 0 & 1 & 1\\
0 & 0 & 0 & 1 & 1 & 0 & 1 & 1 & 0 & 0 & 0 & 1 & 1 & 1 & 0 & 0 & 0 & 0 & 1 & 1 & 0 & 1 & 0 & 1 & 1 & 1\\
1 & 0 & 1 & 0 & 1 & 0 & 1 & 1 & 0 & 1 & 0 & 0 & 0 & 0 & 1 & 0 & 1 & 1 & 0 & 0 & 1 & 1 & 0 & 0 & 1 & 1\\
1 & 1 & 1 & 1 & 0 & 0 & 0 & 1 & 0 & 1 & 0 & 0 & 1 & 0 & 1 & 1 & 1 & 1 & 0 & 1 & 0 & 1 & 1 & 1 & 1 & 1\\
0 & 1 & 1 & 1 & 1 & 0 & 1 & 1 & 1 & 0 & 1 & 0 & 1 & 0 & 1 & 0 & 0 & 0 & 1 & 0 & 0 & 0 & 0 & 1 & 1 & 0\\
1 & 0 & 0 & 1 & 1 & 1 & 1 & 1 & 1 & 1 & 1 & 0 & 1 & 1 & 1 & 1 & 1 & 0 & 1 & 1 & 0 & 0 & 1 & 1 & 0 & 1\\
1 & 1 & 1 & 0 & 1 & 1 & 0 & 0 & 1 & 0 & 0 & 1 & 1 & 0 & 1 & 0 & 0 & 0 & 0 & 0 & 1 & 1 & 0 & 0 & 0 & 0\\
1 & 1 & 0 & 1 & 0 & 0 & 0 & 0 & 0 & 1 & 1 & 1 & 0 & 0 & 0 & 0 & 1 & 1 & 1 & 0 & 0 & 1 & 1 & 1 & 1 & 0\\
0 & 1 & 1 & 0 & 1 & 0 & 0 & 1 & 0 & 0 & 1 & 1 & 1 & 1 & 1 & 1 & 1 & 1 & 1 & 1 & 0 & 0 & 0 & 1 & 1 & 1\\
1 & 0 & 0 & 1 & 0 & 1 & 0 & 0 & 0 & 0 & 0 & 1 & 1 & 1 & 0 & 0 & 0 & 0 & 1 & 0 & 1 & 1 & 0 & 1 & 0 & 1\\
0 & 1 & 0 & 0 & 1 & 1 & 1 & 0 & 1 & 1 & 1 & 0 & 0 & 0 & 1 & 1 & 0 & 1 & 1 & 0 & 1 & 0 & 0 & 1 & 0 & 0\\
\end{array}\right].
\]
\end{example}

\section{Optimal self-orthogonal codes of dimensions 7 and 8}
In~\cite{Bouyukliev2006}, Bouyukliev et al. found the parameters of distance-optimal binary self-orthogonal codes of dimensions $3$ to $10$ up to length $40$. Shi et al. characterized the explicit parameters of $[n, 7]$ distance-optimal self-orthogonal codes of any length $n$ in~\cite{Shi2025}. Recently, Li et al. determined parameters of $[n, 8]$ distance-optimal self-orthogonal codes of some lengths up to $254$ in~\cite{Li2025}.

\begin{table}[ht]
\centering
\caption{Parameters of $d_{SO}(n, 7)$ up to $n\le 126$}
\label{tab:my_table_7col}
\small
\begin{tabular}{c|r|c|c|c|r|c|c}
\Xhline{2\arrayrulewidth}
\multirow{2}{*}{$n$} & \multicolumn{3}{c|}{$d_{SO}(n, 7)$} & \multirow{2}{*}{$n$} & \multicolumn{3}{c}{$d_{SO}(n, 7)$} \\
\cline{2-4} \cline{6-8}
 & Optimal[ref] & Our result & $No.$ &  & Optimal[ref] & Our result & $No.$  \\
\Xhline{2\arrayrulewidth}
27 & 12~\cite{Bouyukliev2006} & 12 & 1 & 78 & 36~\cite{Shi2025} & 36 & 5 \\\hline
29 & 12~\cite{Bouyukliev2006} & 12 & 4 & 79 & 36~\cite{Shi2025} & 36 & 5 \\\hline
30 & 12~\cite{Bouyukliev2006} & 10 & & 80 & 38~\cite{Shi2025} & 36 & \\\hline
35 & 16~\cite{Bouyukliev2006} & 16 & 1 & 82 & 40~\cite{Shi2025} & 40 & 1 \\\hline
36 & 16~\cite{Bouyukliev2006} & 16 & 1 & 83 & 40~\cite{Shi2025} & 40 & 2 \\\hline
37 & 16~\cite{Bouyukliev2006} & 16 & 4 & 84 & 40~\cite{Shi2025} & 40 & 3 \\\hline
38 & 16~\cite{Bouyukliev2006} & 16 & 5 & 91 & 44~\cite{Shi2025} & 44 & 2 \\\hline
42 & 18~\cite{Shi2025} & 18 & 1 & 92 & 44~\cite{Shi2025} & 44 & 4 \\\hline
43 & 20~\cite{Shi2025} & 20 & 1 & 95 & 46~\cite{Shi2025} & 46 & 1 \\\hline
48 & 22~\cite{Shi2025} & 20 & & 96 & 48~\cite{Shi2025} & 48 & 1 \\\hline
49 & 22~\cite{Shi2025} & 22 & 1 & 105 & 52~\cite{Shi2025} & 52 & 1 \\\hline
50 & 24~\cite{Shi2025} & 24 & 1 & 106 & 52~\cite{Shi2025} & 50 & \\\hline
51 & 24~\cite{Shi2025} & 24 & 1 & 107 & 52~\cite{Shi2025} & 52 & 1 \\\hline
59 & 28~\cite{Shi2025} & 28 & 1 & 111 & 54~\cite{Shi2025} & 54 & 1 \\\hline
61 & 28~\cite{Shi2025} & 28 & 1 & 112 & 56~\cite{Shi2025} & 56 & 1 \\\hline
63 & 28~\cite{Shi2025} & 28 & 32 & 119 & 58~\cite{Shi2025} & 58 & 1 \\\hline
64 & 32~\cite{Shi2025} & 32 & 1 & 120 & 60~\cite{Shi2025} & 60 & 1 \\\hline
75 & 36~\cite{Shi2025} & 36 & 1 & 124 & 60~\cite{Shi2025} & 60 & 1 \\\hline
76 & 36~\cite{Shi2025} & 36 & 1 & 126 & 62~\cite{Shi2025} & 62 & 1 \\
\Xhline{2\arrayrulewidth}
\end{tabular}
\end{table}

In this section, we construct self-orthogonal codes of dimensions $7$ and $8$ by applying our method for producing shortest self-orthogonal embeddings to codes from MAGMA's database of ``best-known linear codes''. For each code, we first find one shortest self-orthogonal embedding and then apply Theorem~\ref{ortho-gene} to examine all shortest self-orthogonal embeddings obtained from it under the action of the orthogonal group.

For the case of dimension $7$, we have obtained distance-optimal self-orthogonal codes in most cases. This result is given in Table~\ref{tab:my_table_7col}. The column labeled `Our result' below $d_{SO}(n, 7)$ in Table~\ref{tab:my_table_7col} represents the highest minimum distance of self-orthogonal codes we have constructed with corresponding lengths and dimensions. The column labeled `$No.$' represents the number of inequivalent distance-optimal self-orthogonal codes among those we have constructed. We have verified that many of the codes we have constructed are equivalent to the codes given in~\cite{Shi2025}. However, since~\cite{Shi2025} gives only one construction for the distance-optimal self-orthogonal codes of each length, the codes we have obtained for length 29, 37, 38, 63, 78, 79, 83, 84, 91, and 92 include codes inequivalent to the code presented in \cite{Shi2025}.

\begin{table}[!htb]
\centering
\caption{Parameters of $d_{SO}(n, 8)$ up to $n\le 254$}
\label{tab:my_table_8col}
\small
\begin{tabular}{c|r|c|c|c|r|c|c}
\Xhline{2\arrayrulewidth}
\multirow{2}{*}{$n$} & \multicolumn{3}{c|}{$d_{SO}(n, 8)$} & \multirow{2}{*}{$n$} & \multicolumn{3}{c}{$d_{SO}(n, 8)$} \\
\cline{2-4} \cline{6-8}
 & Optimal[ref] & Our result & $No.$ &  & Optimal[ref] & Our result & $No.$  \\
\Xhline{2\arrayrulewidth}
29 & 12~\cite{Bouyukliev2006} & 12 & 1 & 145 & 68-70~\cite{Li2025} & 68 & 1\\
\hline
31 & 12~\cite{Bouyukliev2006} & 12 & 2 & 146 & 70~\cite{Li2025} & 68 &\\
\hline
33 & 12~\cite{Bouyukliev2006} & 12 & 288 & 147 & 72~\cite{Li2025} & 72 & 1\\
\hline
36 & 16~\cite{Bouyukliev2006} & 16 & 1 & 149 & 72~\cite{Li2025} & 72 & 2\\
\hline
37 & 16~\cite{Bouyukliev2006} & 16 & 1 & 150 & 72~\cite{Li2025} & 72 & 1\\
\hline
38 & 16~\cite{Bouyukliev2006} & 16 & 4 & 151 & 72~\cite{Li2025} & 72 & 28\\
\hline
43 & 18~\cite{Li2025} & 16 & & 154 & 74~\cite{Li2025} & 74 & 1\\
\hline
45 & 20~\cite{Li2025} & 20 & 1 & 155 & 76~\cite{Li2025} & 76 & 1\\
\hline
48 & 20~\cite{Li2025} & 20 & 6 & 157 & 76~\cite{Li2025} & 76 & 3\\
\hline
49 & 20~\cite{Li2025} & 20 & 10 & 159 & 76~\cite{Li2025} & 76 & 1\\
\hline
50 & 22~\cite{Li2025} & 20 & & 161 & 78~\cite{Li2025} & 78 & 1\\
\hline
51 & 24~\cite{Li2025} & 24 & 1 & 162 & 80~\cite{Li2025} & 80 & 1\\
\hline
61 & 28~\cite{Li2025} & 28 & 1 & 170 & 82~\cite{Li2025} & 82 & 1\\
\hline
62 & 28~\cite{Li2025} & 28 & 1 & 171 & 84~\cite{Li2025} & 84 & 1\\
\hline
64 & 28~\cite{Li2025} & 26 & & 177 & 86~\cite{Li2025} & 86 & 1\\
\hline
65 & 28~\cite{Li2025} & 28 & 33 & 178 & 88~\cite{Li2025} & 88 & 1\\
\hline
67 & 30~\cite{Li2025} & 30 & 1 & 187 & 92~\cite{Li2025} & 92 & 1\\
\hline
68 & 32~\cite{Li2025} & 32 & 2 & 189 & 92~\cite{Li2025} & 92 & 2\\
\hline
69 & 32~\cite{Li2025} & 32 & 1 & 190 & 92~\cite{Li2025} & 92 & 12\\
\hline
77 & 36~\cite{Li2025} & 36 & 1 & 191 & 92-94~\cite{Li2025} & 94* & 1\\
\hline
80 & 36~\cite{Li2025} & 36 & 5 & 192 & 96~\cite{Li2025} & 96 & 1\\
\hline
81 & 36~\cite{Li2025} & 36 & 52 & 203 & 100~\cite{Li2025} & 100 & 1\\
\hline
84 & 40~\cite{Li2025} & 40 & 2 & 205 & 100~\cite{Li2025} & 100 & 2\\
\hline
85 & 40~\cite{Li2025} & 40 & 1 & 206 & 100~\cite{Li2025} & 100 & 5\\
\hline
88 & 40~\cite{Li2025} & 40 & 4 & 207 & 100-102~\cite{Li2025} & 100 & 5\\
\hline
91 & 40-42~\cite{Li2025} & 42* & 1 & 209 & 102~\cite{Li2025} & 102 & 1\\
\hline
92 & 44~\cite{Li2025} & 44 & 2 & 210 & 104~\cite{Li2025} & 104 & 1\\
\hline
98 & 44-46~\cite{Li2025} & 46* & 1 & 211 & 104~\cite{Li2025} & 104 & 1\\
\hline
99 & 48~\cite{Li2025} & 48 & 1 & 212 & 104~\cite{Li2025} & 104 & 1\\
\hline
108 & 52~\cite{Li2025} & 52 & 1 & 216 & 106~\cite{Li2025} & 106 & 2\\
\hline
109 & 52~\cite{Li2025} & 52 & 1 & 217 & 108~\cite{Li2025} & 108 & 1\\
\hline
110 & 52~\cite{Li2025} & 52 & 3 & 223 & 110~\cite{Li2025} & 110 & 2\\
\hline
114 & 52-54~\cite{Li2025} & 54* & 1 & 224 & 112~\cite{Li2025} & 112 & 1\\
\hline
115 & 56~\cite{Li2025} & 56 & 1 & 232 & 114~\cite{Li2025} & 114 & 1\\
\hline
124 & 60~\cite{Li2025} & 60 & 1 & 233 & 116~\cite{Li2025} & 116 & 1\\
\hline
125 & 60~\cite{Li2025} & 60 & 3 & 239 & 118~\cite{Li2025} & 118 & 1\\
\hline
126 & 60~\cite{Li2025} & 60 & 3 & 240 & 120~\cite{Li2025} & 120 & 1\\
\hline
127 & 60~\cite{Li2025} & 60 & 1 & 247 & 122~\cite{Li2025} & 122 & 2\\
\hline
128 & 64~\cite{Li2025} & 64 & 1 & 248 & 124~\cite{Li2025} & 124 & 1\\
\hline
142 & 68~\cite{Li2025} & 66 & & 252 & 124~\cite{Li2025} & 124 & 1\\
\hline
143 & 68~\cite{Li2025} & 68 & 2 & 254 & 126~\cite{Li2025} & 126 & 1\\
\hline
144 & 68~\cite{Li2025} & 68 & 7 &  &  &  & \\
\Xhline{2\arrayrulewidth}
\end{tabular}
\end{table}

Similarly, for the case of dimension 8, we have obtained distance-optimal self-orthogonal codes in most cases.  Our results are given in Table~\ref{tab:my_table_8col}. In the table, asterisk (*) indicates that our construction attains the previously known upper bound and thereby determines the exact value of $d_{SO}(n, 8)$. The column `$No.$' denotes the number of inequivalent self-orthogonal codes we constructed. This count includes codes confirmed to be distance-optimal, as well as those whose parameters are equal to the best-known self-orthogonal codes even if their optimality is not yet known. Moreover, while~\cite{Li2025} presented parameters for distance-optimal self-orthogonal codes of dimension $8$, the minimum distance of distance-optimal self-orthogonal codes for several lengths remained undetermined. Through our construction, we have determined the parameters of distance-optimal self-orthogonal codes for some of these previously open cases. We have obtained $[91, 8, 42],\, [98, 8, 46],\,[114, 8, 54]$, and $[191, 8, 94]$ codes. To the best of our knowledge, these codes were not previously known. We provide the generator matrices for these codes, which are presented in hexadecimal format after appending them with zero columns to ensure the length is a multiple of 4:
\[
G(\mathcal{C}_{[91, 8, 42]})=\left[
\begin{array}{c}
\texttt{9178c93aad6db724a17528e} \\
\texttt{524e3aa98bd1d2a3d2e97c8} \\
\texttt{302ae870277bdc487aa05b4} \\
\texttt{0b12221ce1c887cef9dbcee} \\
\texttt{070e41fc1ff701c1f1073f0} \\
\texttt{00fe7803ff87ffc03e1f02c} \\
\texttt{0001f800007fffffc01ffcc} \\
\texttt{000007ffffffffffffe0006}
\end{array}
\right]
\]

\[
G(\mathcal{C}_{[98, 8, 46]})=\left[
\begin{array}{c}
\texttt{80c090d0a8f8e4ecdcbcdafb8} \\
\texttt{40604868547c72766e5e6d7d0} \\
\texttt{203024342a3e393b372fb6bf8} \\
\texttt{1018121a151f9c9d9b975b5f0} \\
\texttt{080c090d8a8f4ececdcbadae4} \\
\texttt{0406848645c72767e6e5d6d70} \\
\texttt{02034243a2e393b373f26bea4} \\
\texttt{018121a151f1c9d9b979b5f50}
\end{array}
\right]
\]

\[
G(\mathcal{C}_{[114, 8, 54]})=\left[
\begin{array}{c}
\texttt{80e090b0a898f8a4d4f4ecdcbcff8} \\
\texttt{40704858544c7c526a7a766e5e7f0} \\
\texttt{2038242c2a263e29353d3b372fbe4} \\
\texttt{101c121615131f949a9e9d9b97df0} \\
\texttt{080e090b8a898f4a4d4fcecdcbee4} \\
\texttt{0407848545c4c725a6a767e6e5f70} \\
\texttt{028342c2a262e39253d3b373f2fa4} \\
\texttt{01c121615131f149a9e9d9b979fd0}
\end{array}
\right]
\]

\[
G(\mathcal{C}_{[191, 8, 94]})=\left[
\begin{array}{c}
\texttt{80b651765a7a08cb754997e9700e853acce5f301d7b8d1f0} \\
\texttt{412edb24a676e85bece929278b70dcfcce124854b11e71bc} \\
\texttt{210b43ef6f85d10d00d633f55764b8bb8959f8a1faee0188} \\
\texttt{1036f4a81c1885b45aafcb87cfc8f1e8db0d3f83e1f048c6} \\
\texttt{099cc9b65fffbfb4aadb2a19329df8981180900c47bb2ae4} \\
\texttt{05215e997ff8240c8ef265c31041575498e35bba0776d7ec} \\
\texttt{028caea7f532f75d2d15c371119b4272a337bf84ed22c538} \\
\texttt{006106140007bffe9bd3bdef901800896a5abedbffdfd4e0}
\end{array}
\right].
\]
\section{Conclusion}
In this paper, we have obtained the minimum length of a self-orthogonal embedding of a binary linear code $\C$, or equivalently, the minimum number of columns that must be added to a generator matrix of $\C$ to produce a generator matrix for a self-orthogonal code. Let $\C$ be an $[n,k]$ code with $\dim(\Hull(\mathcal{C}))=\ell$. If $\mathcal{C}$ is an even code which is not self-orthogonal, then the length of a shortest self-orthogonal embedding of $\mathcal{C}$ is $n+k-\ell+1$. Otherwise, the length of a shortest self-orthogonal embedding of $\mathcal{C}$ is $n+k-\ell$. We have shown that a shortest self-orthogonal embedding of a Reed-Muller code $\mathcal{R}(r, m)$ has length $2k+1$, and a shortest self-orthogonal embedding of a binary Hamming code $\mathcal{H}_r$ is a self-dual code of length $2k$. We have determined all shortest self-dual embeddings of the Hamming code $\mathcal{H}_4$ up to equivalence, and have found that one of those is equivalent to the shortened Golay code. Using shortest self-orthogonal embeddings, we have obtained many inequivalent distance-optimal self-orthogonal codes of dimensions 7 and 8 across various lengths. Four of these 8-dimensional codes achieve previously unknown parameters.

One of the main problems in coding theory is to find optimal self-orthogonal or self-dual codes with given parameters. Determining the possible minimum distances of the shortest self-orthogonal embedding of a linear code, or constructing an algorithm to produce a shortest self-orthogonal embedding of a code with the largest possible minimum distance, would be interesting topics for further study.

\section*{Acknowledgement}
N. Kaplan was supported by NSF grant DMS 2154223.

J.-L. Kim was supported in part by the BK21 FOUR (Fostering Outstanding Universities for Research) funded by the Ministry of
Education (MOE, Korea) and National Research Foundation of Korea (NRF)
under Grant No. 4120240415042, National Research Foundation of Korea under Grant No. RS-2024-NR121331, and by Basic Science Research Program through the National Research Foundation of Korea (NRF) funded by the Ministry of Science and ICT under Grant No. RS-2025-24534992.

J. Luo and G. Wang are supported by the Project of International Cooperation and Exchanges NSFC (Grant No. 12411540221), National Science Foundation of China (Grant No. 12441102, 12171191,12271199).

\end{document}